\newcommand\nodata{{...} } \newcommand\eg{{\it e.g.} }  \newcommand\etal{et~al.} \newcommand\mdot{M$_{\odot}$~}
 
\newcommand\OVI{\hbox{O~VI}~$\lambda$1035} \newcommand\Lya{Ly$\alpha$}
\newcommand\NV{\hbox{N~V}~$\lambda$1240}
\newcommand\SiIV{\hbox{Si~IV}/\hbox{O~IV}~$\lambda$1402}
\newcommand\CIV{\hbox{C~IV}~$\lambda$1549}
\newcommand\HeII{\hbox{He~II}~$\lambda$1640}
\newcommand\OIIInexttoHe{\hbox{O~III}]~$\lambda$1663}
\newcommand\CIII{\hbox{C~III]}~$\lambda$1909} \newcommand\CII{\hbox{C~$\rm
II$]}~$\lambda$2326} \newcommand\NeIV{\hbox{[Ne~$\rm IV$}]~$\lambda$2424}
\newcommand\MgII{\hbox{Mg~$\rm II$}~$\lambda$2800}
\newcommand\NeV{\hbox{[Ne~$\rm V$]}~$\lambda$3426}
\newcommand\OII{\hbox{[O~II}]~$\lambda$3727}
\newcommand\OIIIblue{\hbox{[O~III}]~$\lambda$4959}
\newcommand\OIII{\hbox{[O~III}]~$\lambda$5007} 
\newcommand\Ha{H$\alpha$} 
 
\newcommand\kms{\ifmmode {\rm\,km\,s^{-1}}\else${\rm\,km\,s^{-1}}$\fi}
 
scaled\magstep1 
\def\spose#1{\hbox to 0pt{#1\hss}}
\newcommand\simlt{\mathrel{\spose{\lower 3pt\hbox{$\mathchar"218$}}
     \raise 2.0pt\hbox{$\mathchar"13C$}}}
\newcommand\simgt{\mathrel{\spose{\lower 3pt\hbox{$\mathchar"218$}}
     \raise 2.0pt\hbox{$\mathchar"13E$}}} \newcommand\aap{{\em A\&A}}
\newcommand\aaps{{\em A\&AS}} \newcommand\aj{{\em AJ}}
\newcommand\apj{{\em ApJ}} \newcommand\apjl{{\em ApJ}}
\newcommand\apjs{{\em ApJS}} 
 
\newcommand\mnras{{\em MNRAS}} \newcommand\nat{{\em Nature}}
 \newcommand\pasp{{\em PASP}}

\documentclass[usenatbib]{mn2e}

\usepackage{psfig}
\usepackage{rotating}
\usepackage{amssymb}

\title[Distant radio galaxies from SUMSS and NVSS]{A search for
distant radio galaxies from SUMSS and NVSS: II. Optical
Spectroscopy\thanks{Based on observations obtained with the
European Southern Observatory, Paranal and La Silla, Chile
(Programs 72.A-0259 and 073.A-0348) and the Australian National
University 2.3m telescope}.}
\author[C. De Breuck \etal]{
\parbox[t]{\textwidth}{
Carlos De Breuck$^1$, Ilana Klamer$^2$, Helen
Johnston$^2$, Richard W.\ Hunstead$^2$, Julia Bryant$^2$, Brigitte
Rocca--Volmerange$^3$, Elaine M.\ Sadler$^2$} \vspace*{6pt} \\
$^1$ European Southern Observatory, Karl Schwarzschild Stra\ss e 2, D-85748 Garching, Germany.\\
$^2$ School of Physics, University of Sydney, Sydney NSW 2006, Australia.\\
$^3$ Institut d'Astrophysique de Paris, UMR7095 CNRS, Universit\'e 
Pierre \& Marie Curie, 98 bis Boulevard Arago, 75014 Paris, France.
}
\pubyear{2005}
\begin{document}
\maketitle

\begin{abstract} 

This is the second in a series of papers presenting observations and
results for a sample of 76 ultra-steep-spectrum (USS) radio sources
in the southern hemisphere designed to find galaxies at high
redshift. Here we focus on the optical spectroscopy program for 53
galaxies in the sample. We report 35 spectroscopic redshifts, based
on observations with the Very Large Telescope (VLT), the New
Technology Telescope (NTT) and the Australian National University's
2.3m telescope; they include five radio galaxies with $z>3$.  
Spectroscopic redshifts for the remaining 18 galaxies could not be
confirmed: three are occulted by Galactic stars, eight show
continuum emission but no discernible spectral lines, whilst the
remaining seven galaxies are undetected in medium-deep VLT
integrations. The latter are either at very high redshift
($z\simgt7$) or heavily obscured by dust.  A discussion of the
efficiency of the USS technique is presented. Based on the similar
space density of $z>3$ radio galaxies in our sample compared with
other USS-selected samples, we argue that USS selection at
843--1400\,MHz is an efficient and reliable technique for finding
distant radio galaxies.

\end{abstract}

\begin{keywords} 

surveys -- radio continuum: general -- radio continuum: galaxies --
galaxies: active -- galaxies: high-redshift

\end{keywords}

\section{Introduction}

A key requirement of structure formation models is being able to predict
the abundance of massive galaxies at high redshift. This has been a
problem for hierarchical galaxy formation models within the
framework of a Lambda cold dark matter ($\Lambda$CDM) cosmogony
\citep[\eg][]{whi91}. These models predict that massive systems assemble
via mergers and coalescence of smaller, intermediate-mass discs
\citep[\eg][]{col91, kau93, bau96, kau99a}.
However, the predicted decline in abundance of early-type galaxies
above $z=1.5$ \citep{kau99b,dev00}, which is a crucial test of these
models, fails to reproduce the observations at high redshift. The
abundance of massive ($\sim10^{12}$~\mdot) sub-millimetre galaxies
with high star-formation rates ($\sim10^3$~\mdot~yr$^{-1}$) at
$z>2$ \citep{ivi02,cha03} is at odds with hierarchical formation.
So too is the existence of massive old elliptical galaxies beyond
$z=1.5$ which have been selected in the radio or near-IR
\citep{dun96,dad00,cim04}, and the massive \citep[M$_{\rm
baryonic}=10^{12}$\mdot;][]{roc04} hosts of $z>3$ radio galaxies
containing large reservoirs ($\sim10^{11}$~\mdot) of molecular gas
and dust \citep{ste03,pap00,deb03,kla05}.

These observations have led to renewed consideration of alternative galaxy
formation models, such as the `down-sizing' idea where the most massive
galaxies form earliest in the Universe, and star formation activity is
progressively shifted to smaller systems
\citep[\eg][]{cow96,kod04,treu05}. They have also prompted revisions of
hierarchical formation models by adding new ingredients and assumptions.
Many models invoke feedback by supernovae, starbursts or active galactic
nuclei (AGN) \citep[\eg][]{gra04,cro05}, while others abandon a universal
initial mass function and consider a full radiative treatment of dust
\citep{bau05}. Several authors have also emphasised the importance of
feedback due to powerful radio jets, which may push back and heat the
ionized gas, reducing or even stopping the cooling flows building up the
galaxy \citep{raw04,cro05}. On the other hand, radio jets could also
induce star formation \citep{rees89,fra04,kla04}, as observed in at least one
high redshift radio galaxy \citep{dey97,bic00}.

It is clear that observations of a large number of high redshift galaxies,
in particular the most massive ones, are essential in order to constrain
the structure formation models described above. Radio galaxies are known
to be among the most massive galaxies known at each redshift
\citep[\eg][]{deb02,roc04}, and are ideal laboratories for studying the
`radio feedback' mechanism described above. Modern radio surveys cover
almost the entire sky, and allow us to pinpoint the most extreme (\eg\ the
most massive) galaxies, provided one can isolate them in radio catalogues
containing up to 2 million sources. We have started such a search for
distant radio galaxies in the southern hemisphere using the 843\,MHz Sydney
University Molonglo Sky Survey \citep[SUMSS;][]{bock99} and the NRAO VLA
Sky Survey \citep[NVSS;][]{con98}. In the first paper of this series
\citep[paper~I;][]{deb04}, we define a sample of 76 high redshift radio
galaxy candidates selected on the basis of their ultra-steep radio
spectra (USS; $\alpha \leq -1.3$, $S_{\nu} \propto \nu^{\alpha}$),
which has been almost the sole way to find $z>3$ radio galaxies
\citep[\eg][]{rot97,deb01,jar01}. We also present high-resolution
radio imaging to obtain accurate positions and morphological
information, and near-IR identifications of the host galaxies. In this
paper~II, we present the results to date of optical spectroscopy to
determine their redshifts. Multi-frequency radio observations and a
discussion of the physics of ultra-steep-spectrum radio galaxies will
be presented in a future paper (Klamer et al. in prep.).  Throughout
this paper, we adopt a flat $\Lambda$CDM cosmology with
H$_0=71$\,km\,s$^{-1}$\,Mpc$^{-1}$, $\Omega_{\rm M}=0.27$ and
$\Omega_{\Lambda}=0.73$ \citep{spe03}.

\section{Observations and data reduction}

We used four different imaging spectrographs for the follow-up
observations of the sources in our USS sample. For identifications
detected on the digitised sky surveys or in the 2 Micron All Sky Survey
\citep[2MASS;][]{cut03}, we used the Australian National University's 2.3m
telescope at Siding Spring Observatory, Coonabarabran, NSW with the Double
Beam Spectrograph \citep[DBS;][]{rod88}. For the remaining sources with
$K\simlt 19$ identifications, we initially attempted to measure redshifts
with the ESO Multi-Mode Instrument (EMMI) on the New Technology Telescope
(NTT). If no redshift could be determined, they were re-observed with the
ESO Very Large Telescope (VLT), along with the faintest $K\simgt 19$
identifications, using the FOcal Reducer and Spectrographs \citep[FORS;
][]{app97} on two of the unit telescopes.

We selected all our targets from the list of 76 USS sources defined in
paper~I. Our goal was to obtain redshift information for the entire
sample, but due to adverse weather conditions (mainly at the NTT), we
could only observe 53 sources. Within the available RA range, we gave
priority to the 53 sources in the sample with $\alpha<-1.3$. In total,
we have observed 41 out of 53 sources with $\alpha<-1.3$ and 12 out of
23 sources with $\alpha>-1.3$. Our spectroscopic sample is thus 77\%
complete for the most important subset of $\alpha<-1.3$ sources.

\subsection{$I-$band imaging} Before attempting optical spectroscopy of
the 10 faintest $K-$band sources, we first imaged these fields in $I-$band
using FORS2 on the VLT (see Table~\ref{imagingjournal}). We split the
observations in exposures of typically 3\,minutes each, while dithering
the frames by a few arcseconds to ensure the object did not fall on a bad
pixel on the detector. The pixel scale of FORS2 is 0\farcs25/pix, and the
seeing during the observations varied between 0\farcs5 and 1\farcs0. We
used the standard imaging reduction steps in {\it IRAF}, consisting of
bias subtraction, flatfielding, and registration of the dithered frames.
We fine-tuned the astrometry using all non-saturated 2MASS stars
(typically $\sim$10 per field) present in the images, yielding a solution
which we estimate to be accurate up to $\sim$0\farcs3, which is sufficient
to identify the host galaxies of the radio sources. Finally, we measured
the magnitudes in 4\arcsec\-diameter apertures using the {\it IRAF} task
{\tt phot}.
                                                                                
\begin{table}
\caption{Journal of the FORS2 $I-$band imaging (4\arcsec\ apertures)}
\label{imagingjournal}
\begin{center}
\begin{footnotesize}
\begin{tabular}{lrrrr}\hline
\multicolumn{1}{c}{Source} & \multicolumn{1}{c}{$t_{\rm exp}$} & 
 \multicolumn{1}{c}{$I$} &
\multicolumn{1}{c}{$K$} & \multicolumn{1}{c}{Seeing}\\ &
\multicolumn{1}{c}{s} & \multicolumn{1}{c}{mag} &
\multicolumn{1}{c}{mag} & \multicolumn{1}{c}{$''$}
\\
\hline
NVSS~J002415$-$324102 &  600 & 23.7$\pm$0.2 & 19.5$\pm$0.4 & 0.5\\
NVSS~J204420$-$334948 & 1620 & $>$24.8      & $>$21 & 0.8\\
NVSS~J230035$-$363410 & 1500 & $>$25        & 19.8$\pm$0.3 & 0.6\\
NVSS~J230527$-$360534 & 1440 & $>$25        & $>$20.6 & 0.6\\
NVSS~J231144$-$362215 & 1620 & 25.2$\pm$0.4 & 20.2$\pm$0.7 & 0.7\\
NVSS~J231727$-$352606 & 1620 & 24.6$\pm$0.3 & $>$20.6 & 0.8\\
NVSS~J232100$-$360223 &  480 & 23.7$\pm$0.3 & 20.0$\pm$0.4  & 0.6\\
NVSS~J232219$-$355816 & 1440 & $>$25        & $>$20.6  & 0.8\\
NVSS~J234137$-$342230 & 1440 & $>$25        & $>$21 & 0.6\\
NVSS~J235137$-$362632 & 1620 & $>$25        & $>$20.4 & 1.0\\
\hline
\end{tabular}
\end{footnotesize}
\end{center}
\end{table}

\begin{table*}
\caption{Journal of the spectroscopic observations}
\label{spectroscopyjournal}
\begin{scriptsize}
\begin{center}
\begin{tabular}{lrrrrrrrrrrrr}\hline
\multicolumn{1}{c}{(1)} & \multicolumn{1}{c}{(2)} & \multicolumn{1}{c}{(3)} & \multicolumn{1}{c}{(4)} & 
\multicolumn{1}{c}{(5)} & \multicolumn{1}{c}{(6)} & \multicolumn{1}{c}{(7)} & \multicolumn{1}{c}{(8)} & 
\multicolumn{1}{c}{(9)} & \multicolumn{1}{c}{(10)} & \multicolumn{1}{c}{(11)}\\
\multicolumn{1}{c}{Source} & \multicolumn{1}{c}{$K$(64\,kpc)} & 
\multicolumn{1}{c}{$z$} & 
\multicolumn{1}{c}{ID$^{\dagger}$} & 
\multicolumn{1}{c}{Log($L_3$)$^{**}$} & \multicolumn{1}{c}{Instr.$^{\ddagger}$} & 
\multicolumn{1}{c}{$t_{\rm exp}$} & \multicolumn{1}{c}{Slit PA} & \multicolumn{1}{c}{Extraction} & RA(J2000) & DEC(J2000) \\
 & \multicolumn{1}{c}{mag} &  & & &  & \multicolumn{1}{c}{s} & \multicolumn{1}{c}{\degr} & \multicolumn{1}{c}{\arcsec$\times$\arcsec} & $^h$\quad$^m$\quad$^s$ & \degr\quad \arcmin\quad \arcsec \\
\hline            
NVSS~J001339$-$322445\rlap{*} & 13.01  & 0.2598$\pm$0.0003& RG     &  25.12  & 2dFGRS & survey & \nodata &2\arcsec\ fibre & 00 13 39.15 & $-$32 24 43.9\\
NVSS~J002001$-$333408 & 17.83          & continuum       & \nodata & $<$26.6 & EMMI   &   2400 & 345     & 1.0$\times$1.0 & 00 20 01.14 & $-$33 34 07.2\\
NVSS~J002112$-$321208 & 19.06          & undetected      & \nodata & \nodata & FORS2  &   2400 & 0       & \nodata        & 00 21 12.40 & $-$32 12 10.1\\
                      &                & undetected      & \nodata & \nodata & FORS1  &   1800 & 0       & \nodata        & 00 21 12.56 & $-$32 12 08.6\\
NVSS~J002131$-$342225 & 15.19          & 0.249$\pm$0.001 & RG      &  24.06  & EMMI   &   1800 & 206     & 2.0$\times$2.8 & 00 21 31.31 & $-$34 22 22.6\\
NVSS~J002219$-$360728 & $>$14.12       & 0.364$\pm$0.001 & RG      &  24.43  & DBS    &   4000 & 90      & 2.0$\times$9.1 & 00 22 19.50 & $-$36 07 29.5\\
NVSS~J002402$-$325253 & 18.98          & 2.043$\pm$0.002 & RG      &  26.89  & FORS1  &   1500 & 0       & 1.0$\times$3.2 & 00 24 02.30 & $-$32 52 54.7\\
NVSS~J002415$-$324102 & 18.84          & continuum       & \nodata & $<$27.0 & FORS2  &    600 & 0       & 1.0$\times$1.5 & 00 24 15.09 & $-$32 41 02.4\\
NVSS~J002427$-$325135 & 17.71          & continuum       & \nodata & $<$27.1 & EMMI   &   2400 & 359     & 1.0$\times$4.3 & 00 24 27.72 & $-$32 51 35.5\\
NVSS~J002627$-$323653 & $>$14.12       & 0.43$\pm$0.01   & RG      &  25.01  & DBS    &   4000 & 70      & 2.0$\times$2.7 & 00 26 27.87 & $-$32 36 52.4\\
NVSS~J011032$-$335445 & 18.19          & continuum       & \nodata & $<$26.6 & EMMI   &   4000 & 332     & 1.5$\times$1.2 & 01 10 31.99 & $-$33 54 42.7\\
NVSS~J011606$-$331241 & 18.00          & 0.352$\pm$0.001 & RG      &  24.49  & FORS1  &   1200 & 0       & 1.0$\times$2.4 & 01 16 06.76 & $-$33 12 43.0\\
NVSS~J012904$-$324815\rlap{*} & 12.58  & 0.180$\pm$0.001 & RG      &  24.15  & DBS    &   2400 & 235     & 2.0$\times$4.6 & 01 29 04.26 & $-$32 48 13.2\\
                      &                & 0.1802$\pm$0.0003& RG     &  24.15  & 2dFGRS & survey & \nodata &2\arcsec\ fibre & 01 29 04.29 & $-$32 48 13.7\\
NVSS~J015223$-$333833 & 17.99          & undetected      & \nodata & \nodata & FORS2  &   2160 & 0       & \nodata        & 01 52 22.93 & $-$33 38 36.2\\
NVSS~J015232$-$333952 & 16.19          & 0.618$\pm$0.001 & RG      &  26.23  & DBS    &   2400 & 0       & 2.0$\times$9.1 & 01 52 32.42 & $-$33 39 55.4 \\
NVSS~J015324$-$334117\rlap{*} & 14.12  & 0.1525$\pm$0.0004& RG     &  23.70  & 2dFGRS & survey & \nodata &2\arcsec\ fibre & 01 53 24.96 & $-$33 41 25.4\\
NVSS~J015418$-$330150\rlap{*} & 19.81  & undetected      & \nodata & \nodata & FORS1  &   3600 & 0       & \nodata        & 01 54 18.26 & $-$33 01 51.0\\
NVSS~J015544$-$330633\rlap{*} & 16.94  & 1.048$\pm$0.002 & Q       &  26.03  & EMMI   &   1800 & 206     & 2.0$\times$2.3 & 01 55 44.61 & $-$33 06 34.9\\
NVSS~J021308$-$322338 & 19.43          & 3.976$\pm$0.001 & RG      &  27.50  & FORS1  &   2100 & 0       & 1.0$\times$1.6 & 02 13 07.98 & $-$32 23 39.9\\
NVSS~J021716$-$325121 & 18.42          & 1.384$\pm$0.002 & RG      &  26.11  & FORS2  &   1800 & 0       & 1.0$\times$1.5 & 02 17 15.92 & $-$32 51 21.3\\
NVSS~J030639$-$330432 & 17.78          & 1.201$\pm$0.001 & RG      &  26.03  & EMMI   &    500 & 268     & 2.0$\times$2.6 & 03 06 39.77 & $-$33 04 32.5\\
NVSS~J202026$-$372823 & 18.11          & 1.431$\pm$0.001 & RG      &  26.36  & EMMI   &   3600 & 11      & 1.0$\times$1.7 & 20 20 26.98 & $-$37 28 21.0\\
NVSS~J202140$-$373942 & 15.09          & 0.185$\pm$0.001 & RG      &  23.85  & EMMI   &   1800 & 169     & 1.0$\times$1.0 & 20 21 40.59 & $-$37 39 40.2\\
NVSS~J202518$-$355834 & 18.21          & undetected      & \nodata & \nodata & FORS1  &   2400 & 0, 306  & 1.0$\times$1.6 & 20 25 18.38 & $-$35 58 32.3\\
NVSS~J202856$-$353709\rlap{*} & 16.58  & 0.000           & S       & \nodata & EMMI   &   3600 & 169     & 2.0$\times$1.8 & 20 28 56.77 & $-$35 37 06.0\\
NVSS~J202945$-$344812\rlap{*} & 17.34  & 1.497$\pm$0.002 & Q       &  26.57  & EMMI   &   2400 & 177     & 2.0$\times$2.1 & 20 29 45.82 & $-$34 48 15.5\\
NVSS~J204147$-$331731 & 16.86          & 0.871$\pm$0.001 & RG      &  25.44  & EMMI   &   3000 & 175     & 1.0$\times$1.1 & 20 41 47.61 & $-$33 17 29.9\\
NVSS~J204420$-$334948 & $>$21          & undetected      & \nodata & \nodata & FORS2  &   3600 & 0       & \nodata        & 20 44 20.83 & $-$33 49 50.6\\
                      &                & undetected      & \nodata & \nodata & FORS1  &   8100 & 0       & \nodata        & 20 44 20.83 & $-$33 49 50.6\\
NVSS~J213510$-$333703 & 18.58          & 2.518$\pm$0.001 & RG      &  26.84  & FORS1  &   3600 & 0       & 1.0$\times$7.2 & 21 35 10.48 & $-$33 37 04.4\\
NVSS~J225719$-$343954 & 16.48          & 0.726$\pm$0.001 & RG      &  25.55  & EMMI   &    900 & 168     & 1.0$\times$1.6 & 22 57 19.63 & $-$34 39 54.6\\
NVSS~J230123$-$364656 & 19.00          & 3.220$\pm$0.002 & RG      &  27.13  & FORS1  &   1800 & 0       & 1.0$\times$1.6 & 23 01 23.54 & $-$36 46 56.1\\
NVSS~J230203$-$340932 & 18.60          & 1.159$\pm$0.001 & RG      &  25.73  & EMMI   &   3200 & 172     & 1.5$\times$2.7 & 23 02 03.00 & $-$34 09 33.5\\
NVSS~J230404$-$372450\rlap{*} & 17.23  & continuum       & \nodata & $<$26.9 & EMMI   &   3600 & 320     & 1.5$\times$0.9 & 23 04 03.87 & $-$37 24 48.0\\
NVSS~J230527$-$360534 & $>$20.6        & undetected      & \nodata & \nodata & FORS2  &   3600 & 0       & \nodata        & 23 05 27.63 & $-$36 05 34.6\\
NVSS~J230822$-$325027 & 18.55          & 0.000           & S       & \nodata & EMMI   &   2000 & 170     & 2.0$\times$0.9 & 23 08 46.72 & $-$33 48 12.3\\
NVSS~J231016$-$363624 & 14.57          & 0.000           & S       & \nodata & DBS    &   2400 & 90      & 2.0$\times$9.1 & 23 10 16.89 & $-$36 36 33.1 \\
NVSS~J231144$-$362215 & 20.08          & 2.531$\pm$0.002 & RG      &  26.72  & FORS2  &   1800 & 0       & 1.0$\times$2.0 & 23 11 45.22 & $-$36 22 15.4\\
NVSS~J231229$-$371324 & 17.53          & continuum       & \nodata & $<$26.7 & EMMI   &   2700 & 269     & 1.5$\times$0.9 & 23 12 29.09 & $-$37 13 25.8\\
NVSS~J231317$-$352133 & 18.69          & undetected      & \nodata & \nodata & EMMI   &   3600 & 184     & \nodata        & 23 13 17.53 & $-$35 21 33.6\\
NVSS~J231338$-$362708 & 19.24          & 1.838$\pm$0.002 & RG      &  26.33  & FORS1  &   7200 & 123     & 1.0$\times$4.0 & 23 13 38.37 & $-$36 27 09.0\\
NVSS~J231341$-$372504\rlap{*} & 16.97  & continuum       & \nodata & $<$26.5 & EMMI   &   2400 & 347     & 1.5$\times$1.3 & 23 13 41.67 & $-$37 25 01.6\\
NVSS~J231357$-$372413 & 16.53          & 1.393$\pm$0.001 & RG      &  26.35  & EMMI   &   1200 & 330     & 1.5$\times$1.1 & 23 13 57.42 & $-$37 24 15.6\\
NVSS~J231402$-$372925 & 18.60          & 3.450$\pm$0.005 & RG      &  27.95  & FORS1  &   8100 & 0       & 1.0$\times$1.6 & 23 14 02.40 & $-$37 29 27.3\\
NVSS~J231519$-$342710 & 17.91          & 0.970$\pm$0.001 & RG      &  25.69  & EMMI   &   5400 & 193     & 1.5$\times$1.3 & 23 15 19.52 & $-$34 27 13.3\\
NVSS~J231727$-$352606 & $>$20.58       & 3.874$\pm$0.002 & RG      &  27.73  & FORS2  &   3600 & 0       & 1.0$\times$2.5 & 23 17 27.41 & $-$35 26 07.1\\
NVSS~J232001$-$363246 & 19.93          & 1.483$\pm$0.002 & RG      &  26.11  & FORS2  &   2400 & 0       & 1.0$\times$1.5 & 23 20 01.27 & $-$36 32 46.5\\
NVSS~J232100$-$360223 & 19.47          & 3.320$\pm$0.005 & RG      &  27.09  & FORS2  &   1300 & 0       & 1.0$\times$2.0 & 23 21 00.64 & $-$36 02 24.8\\
NVSS~J232322$-$345250\rlap{*} & 17.12  & continuum       & \nodata & $<$26.6 & EMMI   &   3600 & 197     & 2.0$\times$0.9 & 23 23 22.95 & $-$34 52 49.0\\
NVSS~J232408$-$353547 & 13.28         & 0.2011$\pm$0.0004& RG      &  23.86  & 2dFGRS & survey & \nodata &2\arcsec\ fibre & 23 24 08.60 & $-$35 35 45.2\\
NVSS~J232602$-$350321\rlap{*} & 14.56  & 0.293$\pm$0.001 & RG      &  24.34  & DBS    &   5400 & 285     & 2.0$\times$3.6 & 23 26 01.67 & $-$35 03 27.5 \\
NVSS~J232651$-$370909 & 19.28          & 2.357$\pm$0.003 & RG      &  26.94  & FORS1  &   1800 & 0       & 1.0$\times$6.0 & 23 26 51.46 & $-$37 09 10.7\\
NVSS~J233558$-$362236\rlap{*} & 16.67  & 0.791$\pm$0.001 & RG      &  25.46  & EMMI   &   3600 & 262     & 2.0$\times$2.3 & 23 35 59.01 & $-$36 22 41.6\\
NVSS~J234145$-$350624 & 16.81          & 0.644$\pm$0.001 & RG      &  27.14  & DBS    &   1500 & 0       & 2.0$\times$9.1 & 23 41 45.85 & $-$35 06 22.2 \\ 
NVSS~J234904$-$362451 & 17.65          & 1.520$\pm$0.003 & Q       &  26.53  & EMMI   &   1800 & 329     & 1.0$\times$1.2 & 23 49 04.26 & $-$36 24 53.3\\
\hline
\end{tabular}
\end{center}
\end{scriptsize}
\begin{flushleft}
$^*$ These sources do not meet our USS criterion of $\alpha \leq -1.3$ (see paper~I)

$\dagger$ RG = Radio Galaxy; Q = Quasar; S = Star.

$^{**}$ $L_3$ is the radio luminosity at 3 GHz in units of W/Hz. For
`continuum' sources, the upper limit is given for $z=2.3$.

$\ddagger$ 2dFGRS = 2-degree-field galaxy redshift survey
\citep[2dFGRS;][]{col01}; EMMI = NTT/EMMI; FORS1 = VLT/FORS1; FORS2 =
VLT/FORS2; DBS = ANU 2.3m/DBS

\end{flushleft}
\end{table*}

\subsection{Spectroscopy}
Table~\ref{spectroscopyjournal} gives a journal of our spectroscopic
observations. If a candidate was observed at different telescopes, we use
only the best quality spectrum.  The columns are:

\begin{description}
\item[(1)]{Name of the source in IAU J2000 format.}

\item[(2)]{$K-$band magnitude measured with a 64 kpc metric
aperture.  For objects without a redshift we adopted the
$8''$-aperture magnitudes from Paper~I.}

\item[(3)]{Spectroscopic redshift. Sources where no optical emission was
detected over the entire wavelength range are marked by `undetected'.
Sources where continuum emission was detected, but with no discernible
features (emission/absorption lines or clear continuum breaks), are
marked by `continuum'.}

\item[(4)]{The type of object detected (Radio Galaxy, Quasar or Star).}

\item[(5)]{The radio luminosity at a rest-frame frequency of 3\,GHz,
calculated using the 843\,MHz and 1.4\,GHz flux densities given in
paper~I.}

\item[(6)]{The instrument used to obtain the spectrum. Redshifts for three
galaxies were obtained from the 2-degree-field galaxy redshift survey
\citep[2dFGRS;][]{col01}.}

\item[(7)]{The total integration time.}

\item[(8)]{The position angle of the spectroscopic slit on the sky,
measured North through East.}

\item[(9)]{The extraction width of the spectroscopic aperture. For the
2dFGRS, this is the fibre diameter. For all other galaxies, the first
value is the slit width, and the second is the width of the extraction
along the slit.}

\item[(10--11)]{The J2000 coordinates of the object centred in the
spectroscopic slit.}

\end{description}

\subsection{ANU 2.3m}

The 2.3m spectroscopy was carried out on 2003 August 1 to 3. To
maximise the throughput, we replaced the dichroic beamsplitter with a
plane mirror so that all the light was sent into the blue arm of the
spectrograph. The detector was a coated SITe 1752$\times$532 pixel CCD
with a spatial scale of 0$\farcs91$/pix. We used the 158R grating,
providing a dispersion of 4\,\AA/pix and a spectral resolution of
$\sim$10.5\,\AA. The typical useful spectral range is $\sim$4300\,\AA\
to $\sim$9600\,\AA.

\subsection{NTT}

The NTT observations were carried out on 2004 August 10 to 14. To minimise
the effects of differential atmospheric refraction, we observed the
targets with the spectroscopic slit at the parallactic position angle
\citep{fil82}. We used grism~\#2 and a 1\farcs0, 1\farcs5 or 2\farcs0
slit, depending on the seeing, which varied from 0\farcs7 to
2\arcsec. The dispersion was 3.5\,\AA/pix, the spectral resolution
9.5\,\AA, and the spatial resolution 0\farcs33/pix. The typical useful
spectral range is $\sim$4300\,\AA\ to $\sim$9700\,\AA.

\subsection{VLT}

The VLT observations were made in visitor mode on two occasions.  On
2003 November 22 to 25, we used FORS2 on the Unit Telescope 4 Yepun,
while on 2004 August 18 and 19, we used FORS1 on the Unit Telescope 2
Kueyen. Both instruments are similar, but FORS2 has higher sensitivity
in the red. On both instruments, we used the 150I grism and 1\farcs0
slit, providing a dispersion of 5.3\,\AA/pix on FORS1 and 6.7\,\AA/pix
on FORS2. The spectral resolutions are $\sim$23\,\AA\ and 20\,\AA, and
the spatial pixel scales are 0\farcs2/pix and 0\farcs25/pix for FORS1
and FORS2, respectively. The typical useful spectral range is
$\sim$4000\,\AA\ to $\sim$8600\,\AA (FORS1) or 9600\,\AA (FORS2).
To acquire the faint targets into the slit, we used blind
offsets from a nearby ($\simlt$1\arcmin) star in the $K-$band images. As
the FORS instruments have a linear atmospheric dispersion corrector, we
did not need to orient the slit at the parallactic angle. After the first
exposure, we performed a quick data reduction. If the redshift could
already be determined from this first spectrum, the next exposure was
aborted to save observing time. For multiple exposures, we shifted the
individual pointings by 10\arcsec\ along the slit to allow subtraction of
the fringing in the red part of the detector.

\subsection{Data reduction} The same standard data reduction strategy was
used for all four spectrographs. After bias and flatfield correction, we
removed the cosmic rays using the IRAF task {\tt szap}. For those sources
with multiple exposures shifted along the slit, we subtracted each
exposure from the bracketing frame, and removed the residual sky variations
using the IRAF task {\tt background}. Next, we extracted the spectra using
a width appropriate to contain all flux in the extended emission lines
(listed in column 9 of Table~\ref{spectroscopyjournal}). We then
calibrated the one-dimensional spectra in wavelength and flux. Finally, we
corrected the spectra for Galactic reddening using the $E(B-V)$ values
measured from the \citet{schl98} dust maps and the \citet{car89}
extinction law.

\begin{figure*}
\begin{tabular}{ccc}
\vspace{0.2cm}
\psfig{file=NVSSJ002415-324102I.L.PS,width=5.5cm,angle=-90}&
\psfig{file=NVSSJ204420-334948I.C.PS,width=5.5cm,angle=-90}&
\psfig{file=NVSSJ230527-360534I.C.PS,width=5.5cm,angle=-90}\\
\vspace{0.2cm}
\psfig{file=NVSSJ231144-362215I.L.PS,width=5.5cm,angle=-90}&
\psfig{file=NVSSJ231727-352606I.C.PS,width=5.5cm,angle=-90}&
\psfig{file=NVSSJ232100-360223I.C.PS,width=5.5cm,angle=-90}\\
\psfig{file=NVSSJ232219-355816I.C.PS,width=5.5cm,angle=-90}&
\psfig{file=NVSSJ234137-342230I.C.PS,width=5.5cm,angle=-90}&
\psfig{file=NVSSJ235137-362632I.C.PS,width=5.5cm,angle=-90}\\
\end{tabular}
\begin{tabular}{cccc}
\psfig{file=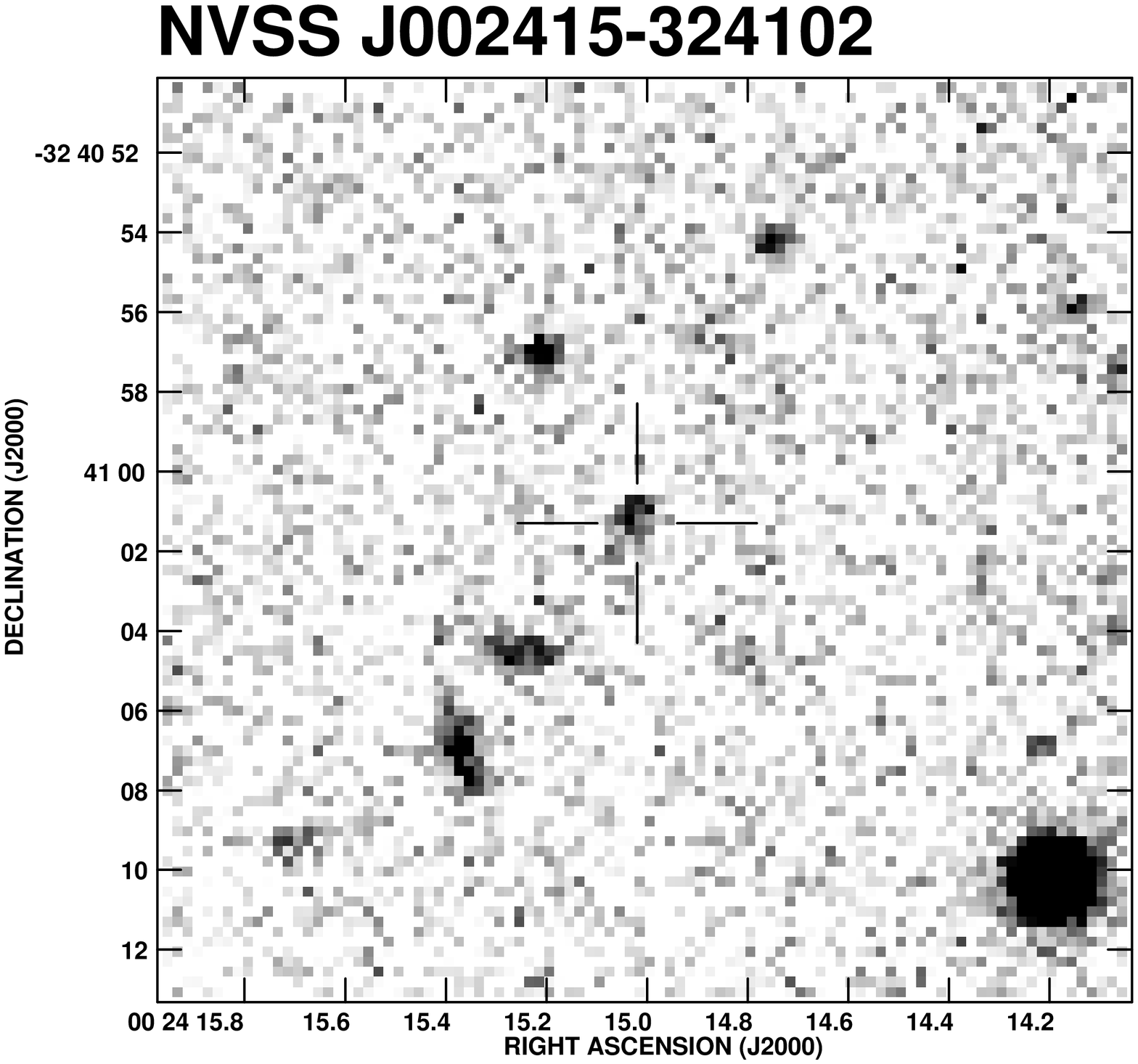,width=4cm,angle=-90}&
\psfig{file=NVSSJ231144-362215I.ZOOM.PS,width=4cm,angle=-90}&
\psfig{file=NVSSJ231727-352606I.ZOOM.PS,width=4cm,angle=-90}&
\psfig{file=NVSSJ232100-360223I.ZOOM.PS,width=4cm,angle=-90}\\
\end{tabular}
\caption{VLT/FORS2 $I-$band images with ATCA 1.4 or 4.8\,GHz contours overlaid. The contour scheme is a geometric progression in
$\sqrt 2$. The first contour level is at 3$\sigma$, where $\sigma$ is the rms noise measured around the sources, as indicated above
each plot. The synthesised beam is shown in the bottom right corner. The $I-$band image of NVSS~J230035$-$363410 is shown in
Fig.~\ref{NVSS2300K.C}. The bottom row shows blow-ups of the four detections.
}
\label{Iimages}
\end{figure*}

\section{Results}
\subsection{$I-$band imaging}

Table~\ref{imagingjournal} lists the results of our FORS2/VLT $I-$band
imaging. Figure~\ref{Iimages} shows the $I-$band images with Australia
Telescope Compact Array 1.4 or 4.8\,GHz radio contours \citep[][Klamer
\etal, in prep.]{deb04} overlaid. We detect four of the ten sources to
a limiting magnitude of $I\simlt 25$. Five out of the six sources with
$I>25$ also remain undetected at $K-$band to a limit of $K>20$.
NVSS~J230035$-$363410 is the only source with a $K-$band
identification which remains undetected in $I-$band down to
$I=25$. This source is discussed further in \S 4.

\subsection{Spectroscopy} 

Tables~\ref{spectroscopyjournal} and \ref{lineparameters} give the
results of our spectroscopy program. To date, we have observed 53 of
the 76 sources in our parent sample, yielding 35 spectroscopic
redshifts, shown in Figure~\ref{1Dspectra} (except the three stars
mentioned below).  Table~\ref{lineparameters} lists the measured
parameters of the emission lines in these spectra, derived using the
procedures described by
\citet{deb01}. In some cases the uncertainties quoted are comparable to
the values themselves; this is due to a combination of low signal-to-noise
data and poorly determined continuum levels.  Nevertheless, we include
these measurements because they confirm the redshifts, and the lines
appear real in the two-dimensional spectra.

Three sources show stellar spectra. Given the low sky density of radio
stars \citep{hel99}, and their extended radio structure, these are almost
certainly foreground stars. The brightness of these stars completely
outshines the background host galaxy, and we therefore exclude them from
further analysis of our sample. This has no effect on the statistics of
our sample.

\begin{figure}
\psfig{file=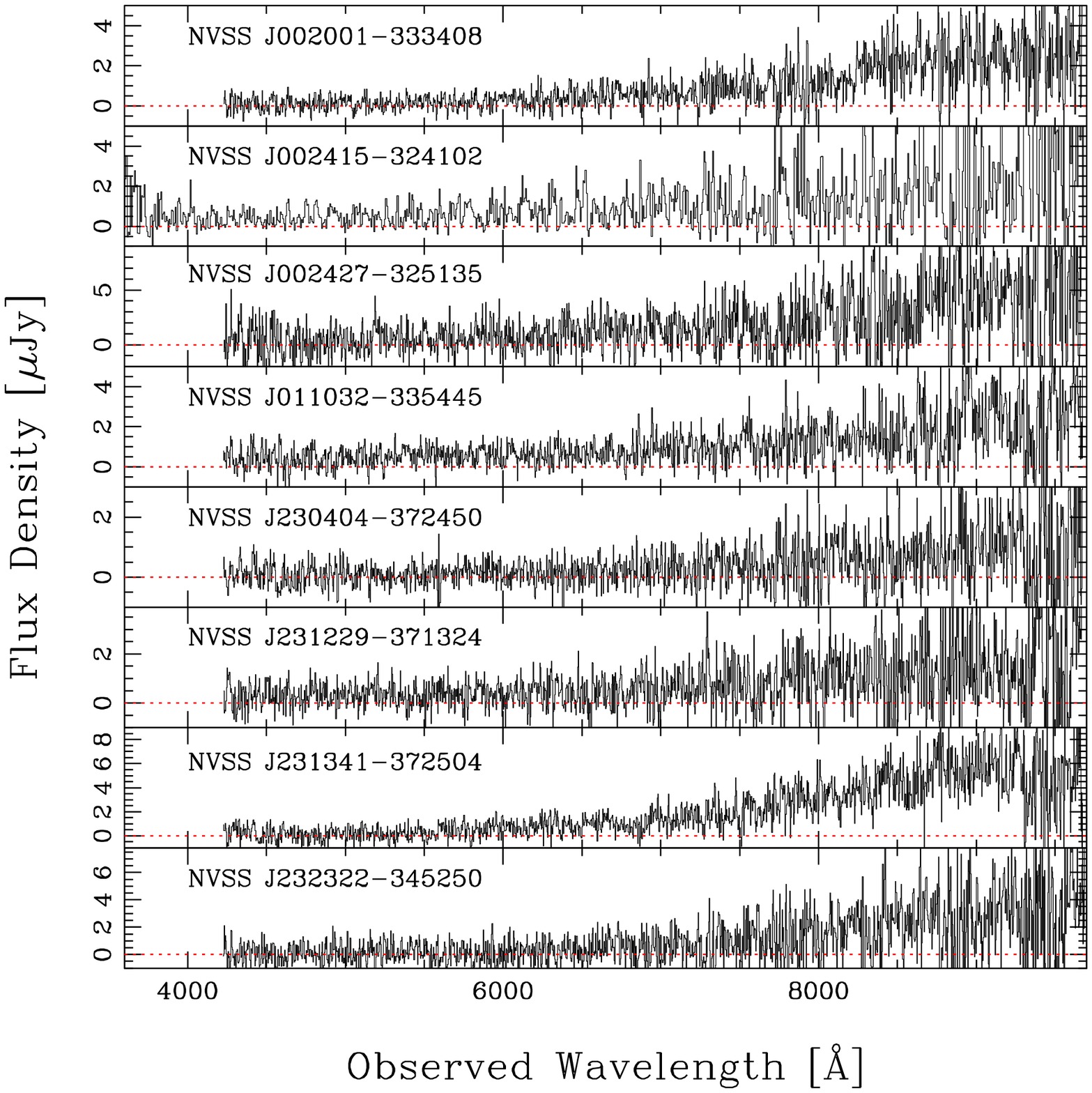,width=8.8cm}
\caption{Spectra obtained with NTT/EMMI or VLT/FORS2 
(NVSS~J002415$-$324102 only) 
showing continuum emission but no identifiable
features.
}
\label{continua}
\end{figure}

There were 15 sources where we could not determine a spectroscopic
redshift.  They fell into two classes: (i) seven sources where we detected
no line or continuum emission over the entire wavelength range (marked
`undetected' in column 2 of Table~\ref{spectroscopyjournal}), and (ii)  
eight sources with continuum emission, but no discernible emission or
absorption lines (marked `continuum' in column 2 of
Table~\ref{spectroscopyjournal} and shown in Fig.~\ref{continua}). The
nature of these sources is discussed in \S 5.1;  they constitute 29\% of
our sample, a fraction very similar to the 35\% in the USS sample of
\citet{deb01}.

\begin{table*}
\caption{Emission line measurements}
\label{lineparameters}
\begin{center}
\begin{tabular}{lrrrrrr}\hline
\multicolumn{1}{c}{Source} & \multicolumn{1}{c}{$z$} & \multicolumn{1}{c}{Line} & \multicolumn{1}{c}{$\lambda_{\rm 
obs}$} & 
\multicolumn{1}{c}{Flux} & \multicolumn{1}{c}{$\Delta v_{\rm FWHM}$} & \multicolumn{1}{c}{$W_{\lambda}^{\rm 
rest}$}\\[1mm]
 &  &  & \multicolumn{1}{c}{\AA} & $10^{-16}$ erg\,s$^{-1}$\,cm$^{-2}$ & km s$^{-1}$ & \multicolumn{1}{c}{\AA} \\
\hline
NVSS~J002219$-$360728 & 0.364$\pm$0.001 & \OII & 5083$\pm$3 & 10.2$\pm$2.0 & 550$\pm$400 & 25$\pm$6 \\
                      &                 & \OIII & 6833$\pm$3 & 5.7$\pm$1.0 & 700$\pm$300 & 15$\pm$3 \\
                      &                 & \Ha   & 8970$\pm$7 & 17.6$\pm$2.3 & 1600$\pm$500 & 100$\pm$20 \\
NVSS~J002402$-$325253 & 2.043$\pm$0.002 & \Lya & 3700$\pm$1 & 4.4$\pm$0.4 & 1150$\pm$620 & 130$\pm$20 \\
                      &                 &   \NV & 3778$\pm$ 1 & 0.15$\pm$0.03 & $<$600        &  5$\pm$1 \\
                      &                 & \SiIV & 4274$\pm$24 & 0.12$\pm$0.06 & $<$2100       &  $<$4    \\
                      &                 &  \CIV & 4714$\pm$ 2 & 1.03$\pm$0.11 &  940$\pm$500  & 51$\pm$7 \\
                      &                 & \HeII & 4992$\pm$ 1 & 1.05$\pm$0.11 & $<$500        & 53$\pm$7 \\
                      &                 & \OIIInexttoHe & 5073$\pm$ 5 & 0.10$\pm$0.03 & $<$7100       &  6$\pm$2 \\
                      &                 & \CIII & 5807$\pm$ 1 & 0.87$\pm$0.09 & $<$410        & 54$\pm$7 \\
                      &                 &  \CII & 7074$\pm$14 & 0.48$\pm$0.08 & 2500$\pm$1100 & 39$\pm$8 \\
                      &                 & \NeIV & 7375$\pm$ 2 & 0.37$\pm$0.05 & $<$340        & 31$\pm$7 \\
                      &                 & \MgII & 8516$\pm$ 3 & 0.31$\pm$0.06 & $<$320        & 23$\pm$7 \\
NVSS~J011606$-$331241 & 0.352$\pm$0.001 & \OII & 5036$\pm$2 & 0.08$\pm$0.03 & $<$460      & $<$13 \\
                      &                 & \OIII & 6769$\pm$3 & 0.47$\pm$0.07 & $<$430      & 100$\pm$25 \\
                      &                 &   \Ha & 8881$\pm$5 & 0.72$\pm$0.12 & 730$\pm$460 & 120$\pm$40 \\
NVSS~J015232$-$333952 & 0.618$\pm$0.001 & \OII & 6030$\pm$2 & 18.4$\pm$2.2 & 700$\pm$225 & 50$\pm$10 \\
                      &                 & \OIIIblue & 8023$\pm$2 & 8.8$\pm$1.2 & 280$\pm$180 & 40$\pm$10 \\
                      &                 & \OIII & 8101$\pm$1 & 27.6$\pm$2.9 & 800$\pm$160 & 130$\pm$20 \\
NVSS~J015544$-$330633 & 1.048$\pm$0.002 & \MgII & 5739$\pm$16 & 20.1$\pm$2.1 & 10400$\pm$1900 & 65$\pm$7 \\
                      &                 & \NeV  & 7016$\pm$6  &  1.3$\pm$0.2 &  1300$\pm$600  &  6$\pm$1 \\
NVSS~J021308$-$322338 &  3.976$\pm$0.001 & \Lya & 6051$\pm$2 & 0.16$\pm$0.02 & $<$400 & 40$\pm$15 \\
NVSS~J021716$-$325121 & 1.384$\pm$0.002 & \OII & 8883$\pm$1 & 1.29$\pm$0.15 & 600$\pm$130 & 150$\pm$50 \\ 
NVSS~J030639$-$330432 & 1.201$\pm$0.001 & \OII & 8203$\pm$2 & 11.2$\pm$1.3 & 800$\pm$200 & 21$\pm$3 \\ 
NVSS~J202026$-$372823 & 1.431$\pm$0.001 & \CIII &  4637$\pm$1 & 0.34$\pm$0.06 & $<$230 & 11$\pm$2 \\
                      &                 & \OII  &  9060$\pm$1 & 3.30$\pm$0.34 & 700$\pm$120 & 110$\pm$20 \\
NVSS~J202945$-$344812 & 1.497$\pm$0.002 & \CIII & 4762$\pm$3 & 18.2$\pm$1.9 & 4100$\pm$540 & 23$\pm$2\\
                      &                 & \NeIV & 6060$\pm$7 &  1.0$\pm$0.2 & 1350$\pm$650 &  2$\pm$0\\
                      &                 & \MgII & 6996$\pm$6 & 14.1$\pm$1.5 & 3900$\pm$600 & 26$\pm$3\\
NVSS~J213510$-$333703 & 2.518$\pm$0.001 & \Lya & 4278$\pm$2 & 1.23$\pm$0.13 & 1350$\pm$600 & $>$518 \\
NVSS~J225719$-$343954 & 0.726$\pm$0.001 & \OII      & 6432$\pm$1 & 2.91$\pm$0.30 & 710$\pm$160 & 92$\pm$12 \\
                      &                 & \OIIIblue & 8553$\pm$2 & 0.99$\pm$0.15 & 520$\pm$220 & 12$\pm$2 \\
                      &                 & \OIII     & 8640$\pm$1 & 2.75$\pm$0.30 & 650$\pm$140 & 33$\pm$4 \\
NVSS~J230123$-$364656 & 3.220$\pm$0.002 & \Lya & 5131$\pm$2 & 0.41$\pm$0.05 & 1200$\pm$500 & 80$\pm$20 \\
NVSS~J230203$-$340932 & 1.159$\pm$0.001 & \OII & 8047$\pm$2 & 0.54$\pm$0.06 & 920$\pm$250 & 33$\pm$5 \\
NVSS~J231144$-$362215 & 2.531$\pm$0.002 & \Lya & 4294$\pm$2 & 0.97$\pm$0.12 & 1600$\pm$400 & \nodata \\ 
NVSS~J231338$-$362708 & 1.838$\pm$0.002 & \Lya & 3462$\pm$11 & 2.14$\pm$0.46 & 3100$\pm$1900 & 35$\pm$10 \\
                      &                 & \CIII & 5417$\pm$4 & 0.10$\pm$0.02 & $<$500 & 3$\pm$1 \\
                      &                 & \CII & 6596$\pm$5 & 0.15$\pm$0.07 & $<$700 & 7$\pm$3 \\
NVSS~J231357$-$372413 & 1.393$\pm$0.001 & \OII & 8920$\pm$2 & 2.15$\pm$0.24 & 910$\pm$180 & 26$\pm$3 \\
NVSS~J231402$-$372925 & 3.450$\pm$0.005 & \Lya & 5411$\pm$6 & 0.17$\pm$0.02 & 1850$\pm$750 & 40$\pm$10 \\
NVSS~J231727$-$352606 & 3.874$\pm$0.002 & \OVI & 5033$\pm$15 & 0.17$\pm$0.19 & 3100$\pm$2000 & 22$\pm$10 \\ 
 & & \Lya & 5930$\pm$7 & 0.43$\pm$0.05 & 3000$\pm$800 & 75$\pm$20 \\
 & & \CIV & 7551$\pm$3 & 0.61$\pm$0.07 & 1250$\pm$250 & 45$\pm$10 \\
 & & \HeII & 7991$\pm$21 & 0.33$\pm$0.07 & 2400$\pm$1200 & 32$\pm$10 \\
NVSS~J232001$-$363246 & 1.483$\pm$0.002 & \CIV & 3857$\pm$20 & 0.43$\pm$0.17 & 3200$\pm$2600 & 20$\pm$10 \\
 & & \HeII & 4076$\pm$1  & 0.29$\pm$0.06 &  250$\pm$200  &  25$\pm$10\\
 & & \CIII & 4733$\pm$2  & 0.25$\pm$0.05 &  500$\pm$300  &  20$\pm$4\\
 & & \CII  & 5765$\pm$2  & 0.15$\pm$0.04 &  300$\pm$250  &  12$\pm$3\\
 & & \NeIV & 6019$\pm$11 & 0.05$\pm$0.04 &  650$\pm$350  & \nodata \\
 & & \NeV  & 8503$\pm$26 & 0.48$\pm$0.15 & 2100$\pm$1200 &  25$\pm$9\\
 & & \OII  & 9253$\pm$2  & 2.72$\pm$0.31 &  750$\pm$150  & 150$\pm$44\\
 
NVSS~J232100$-$360223 & 3.320$\pm$0.005 & \Lya & 5258$\pm$4 & 0.54$\pm$0.07 & 2300$\pm$400 & 65$\pm$15 \\
 & & \CIV & 6683$\pm$5 & 0.05$\pm$0.03 & 325$\pm$300 & \nodata \\
 & & \HeII & 7101$\pm$21 & 0.10$\pm$0.06 & 1400$\pm$900 & \nodata \\ 
NVSS~J232651$-$370909 & 2.357$\pm$0.003 & \Lya & 4078$\pm$2 & 3.4$\pm$0.4 & 1000$\pm$600 & 280$\pm$120 \\
                      &                 & \CIV & 5203$\pm$1 & 0.23$\pm$0.02 & $<$440 & 18$\pm$2 \\
                      &                 & \CIII & 6395$\pm$7 & 0.11$\pm$0.04 & $<$750 & 11$\pm$5 \\
NVSS~J234145$-$350624 & 0.644$\pm$0.001 & \OII & 6126$\pm$3 & 13.5$\pm$2.1 & 740$\pm$260 & 60$\pm$15 \\
                      &                 & \OIII & 8238$\pm$3 & 10.4$\pm$1.5 & 800$\pm$225 & 80$\pm$20 \\
NVSS~J234904$-$362451 & 1.520$\pm$0.003 & \CIII & 4800$\pm$ 4 & 10.9$\pm$1.1 & 5500$\pm$ 600 & 26$\pm$3 \\
                      &                 & \CII  & 5871$\pm$13 &  1.5$\pm$0.2 & 3500$\pm$1600 &  4$\pm$1 \\
                      &                 & \MgII & 7049$\pm$ 9 & 16.8$\pm$1.7 & 7450$\pm$ 900 & 55$\pm$6 \\
                      &                 & \OIII & 7890$\pm$22 &  1.0$\pm$0.2 & 2600$\pm$2300 &  4$\pm$1 \\
                      &                 & \NeV  & 8639$\pm$ 3 &  0.9$\pm$0.1 &  800$\pm$ 300 &  4$\pm$1 \\
\hline
\end{tabular}
\end{center}
\end{table*}

\subsection{Photometric Redshifts}

We attempted to determine photometric redshifts for the targets listed as
`continuum' and `undetected' in Table~\ref{spectroscopyjournal}, using the
photometric redshift code {\it Z-PEG} \citep{leb02}. In the cases where
continuum was detected, we tried to fit a continuum to the observed
$K-$band magnitude, and optical magnitudes derived from the calibrated
spectra.  However, the `undetected' sources had no continuum, and
therefore an upper limit to the optical bands was calculated from the
spectra and fit along with the $K-$band magnitude in an attempt to set a
minimum redshift at which the optical emission would be below the noise
limit.

We performed the same analysis for sources which had redshifts
measured from the spectra. The results showed significant
discrepancies between the fitted values from {\it Z-PEG}, and the
redshifts measured from spectral features. In some cases, no
acceptable fit to the observed colours could be found, despite a secure
redshift having been determined from the spectrum. In other cases, the
{\it Z-PEG} fits were very poorly constrained.  Furthermore, there was
a tendency for the fits to congregate around $z\sim2$.

We believe that photometric redshifts could not be found for several
reasons: template mismatch, absence of IR photometry, and/or dust.
First, the active galaxies in our sample may have a significant
contribution from direct or scattered AGN light, especially in the
optical bands.  If the objects are at $z>1$, as expected from their
faint $K-$band magnitudes (see \S 5.2.3), the optical bands trace the
rest-frame UV emission, and may also be boosted by young star
formation associated with the radio jet activity. In both cases our
galaxies will not be well matched to the template galaxies in {\it
  Z-PEG}.  Second, at the likely redshift range of our sources,
$1<z<4$, the Balmer and 4000\,\AA\ discontinuities shift to the
wavelength range $\sim$0.9 to 2.0\,$\mu$m, a range not covered by our
spectra. \citet{wil01b} found plausible photometric redshifts in the
range $1<z<2$\ for seven sources from the 7C Redshift Survey using
$RIJHK$ photometry, illustrating the importance of near-infrared
photometry for these objects.  Third, it is difficult to assess the
amount of dust in these galaxies.  The photometric redshifts were
therefore not included in this paper.

\subsection{Notes on individual sources}

\begin{figure}
\psfig{file=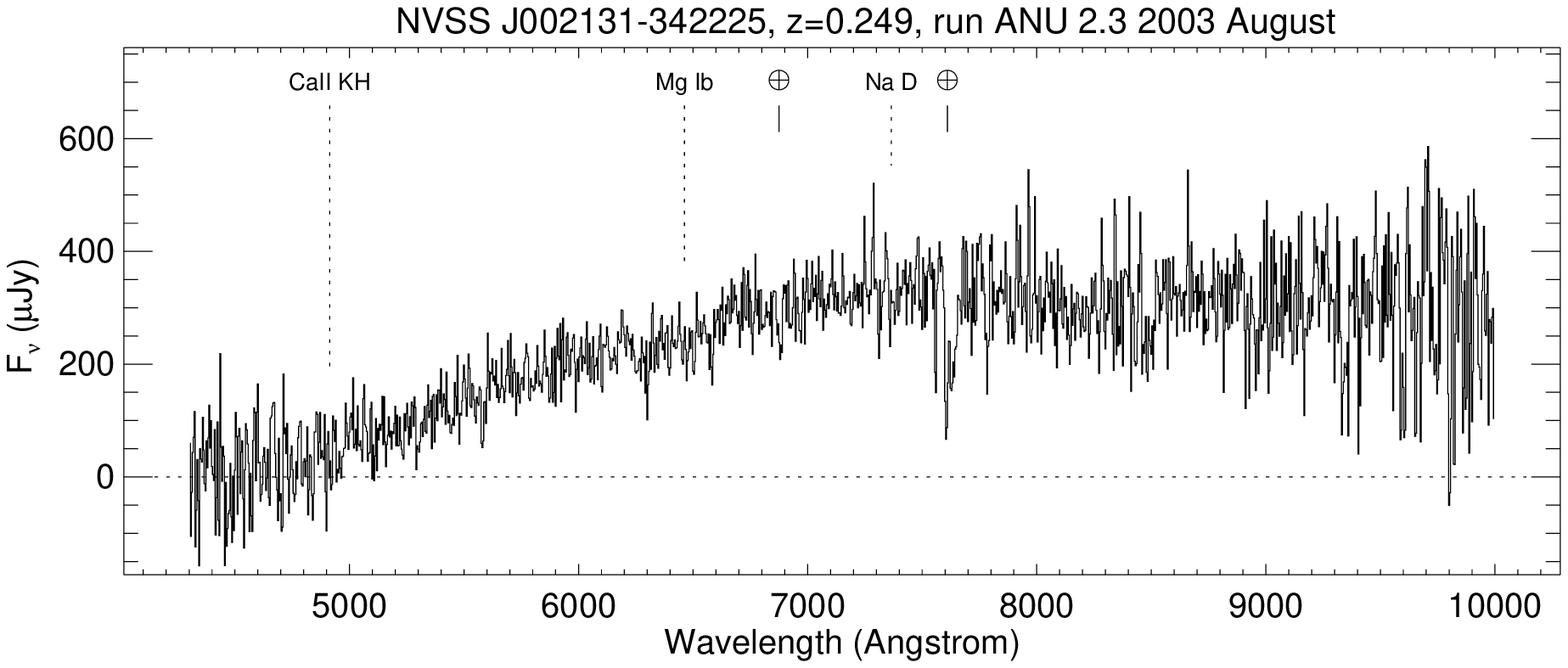,width=8.7cm}
\psfig{file=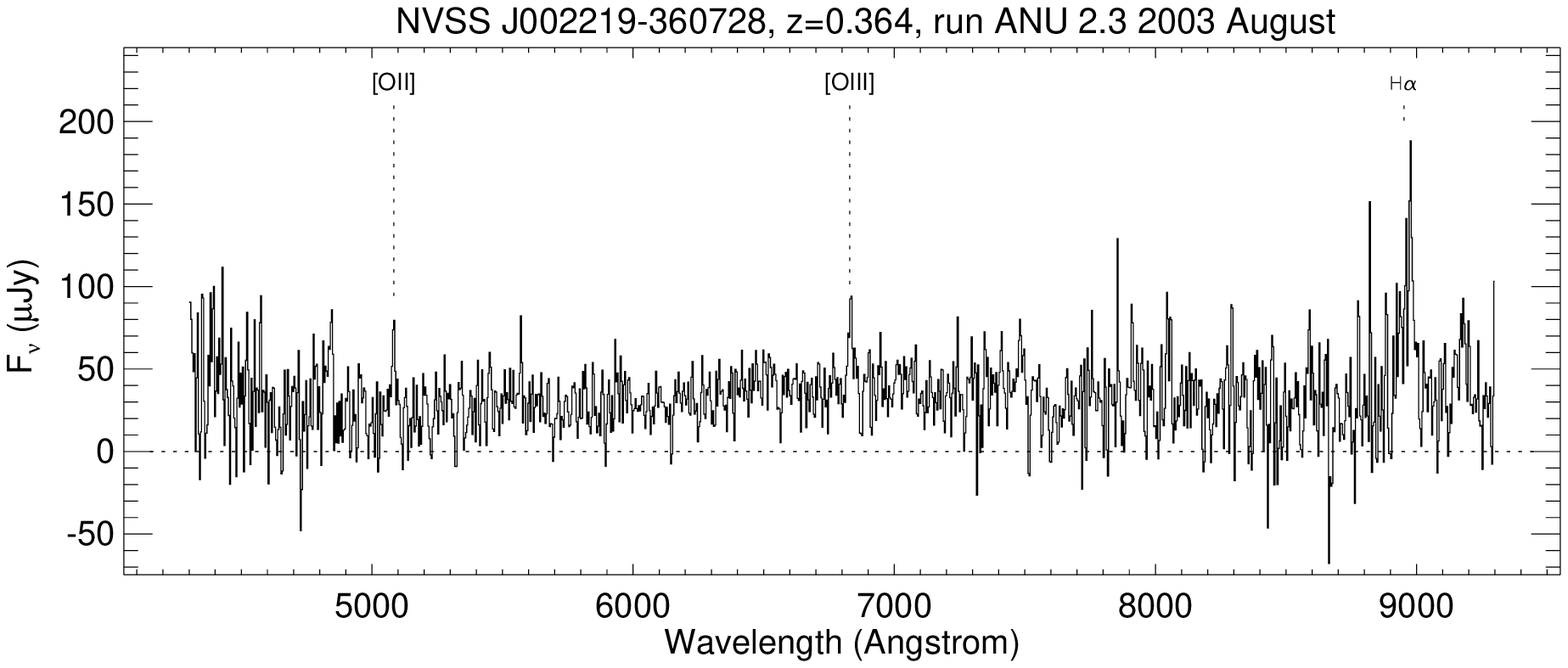,width=8.7cm}
\psfig{file=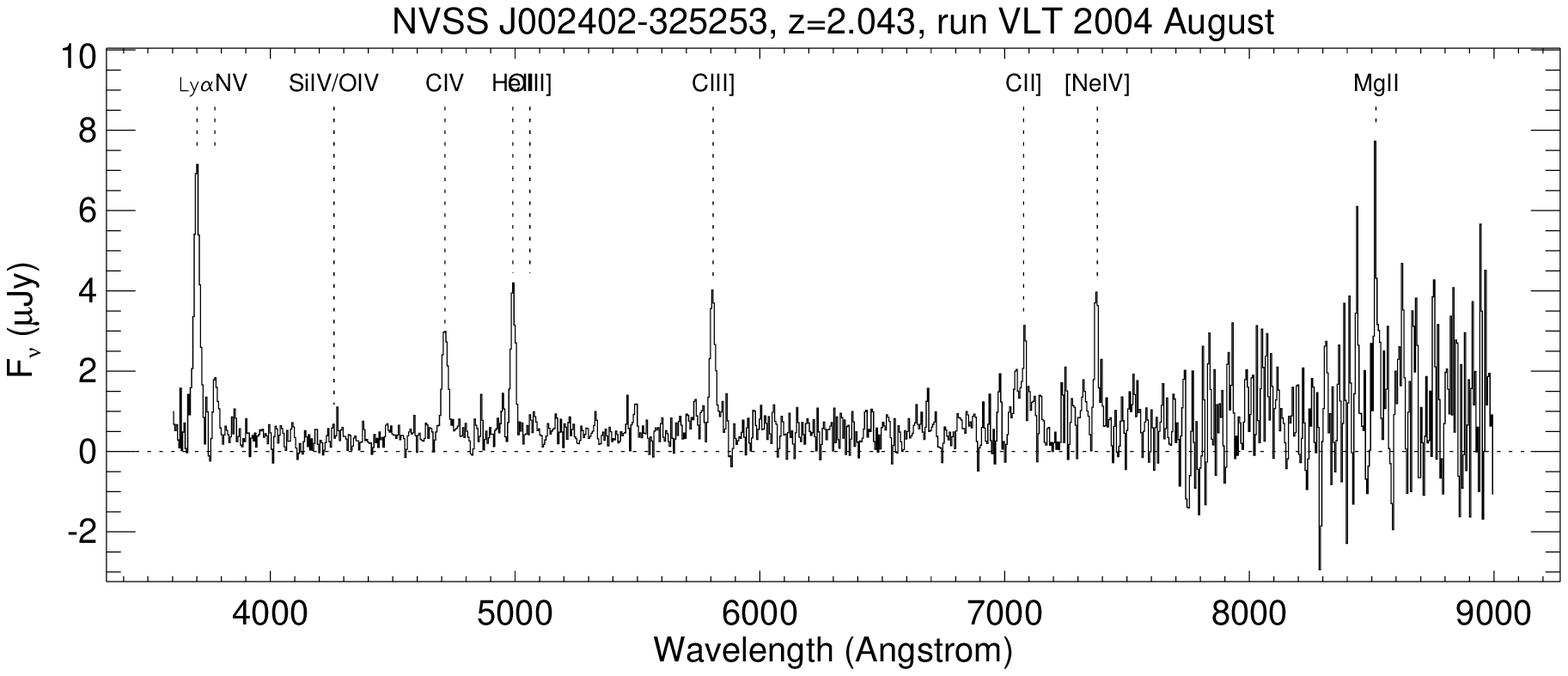,width=8.7cm}
\psfig{file=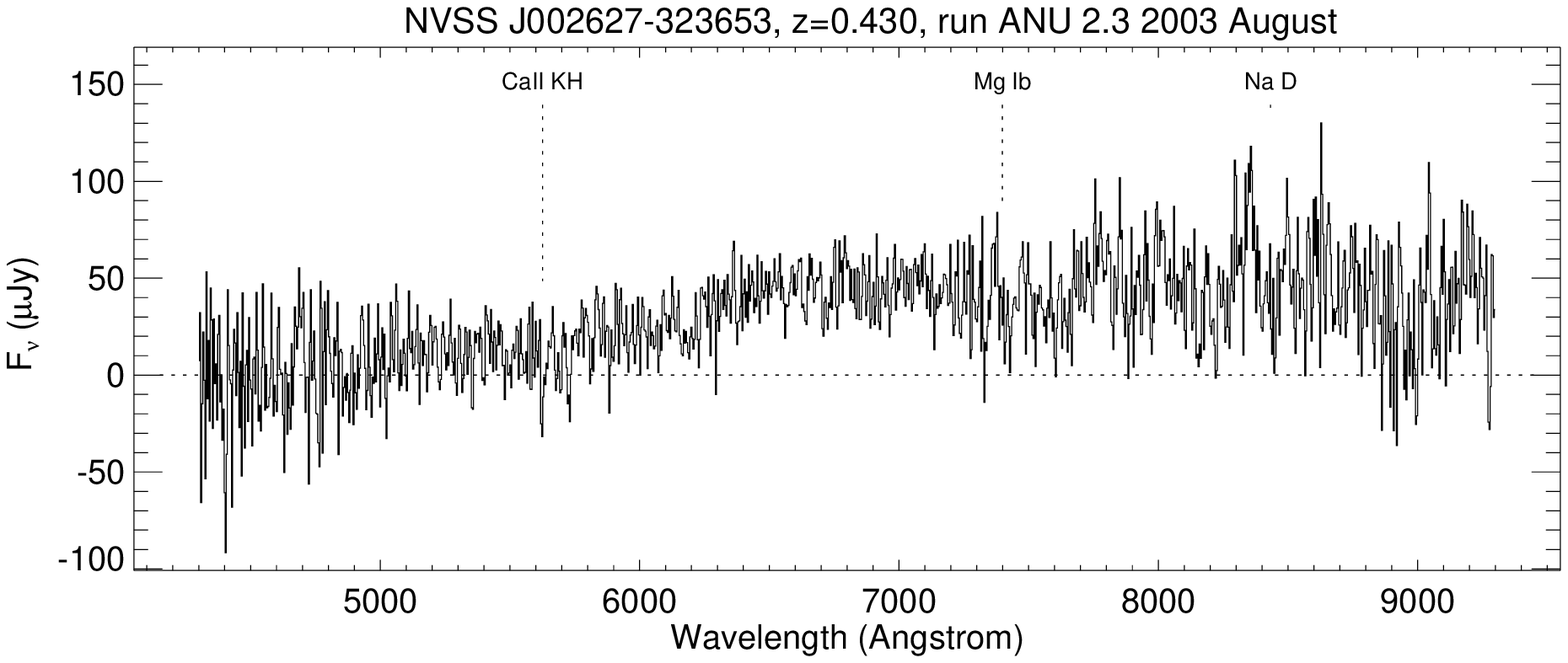,width=8.7cm}
\psfig{file=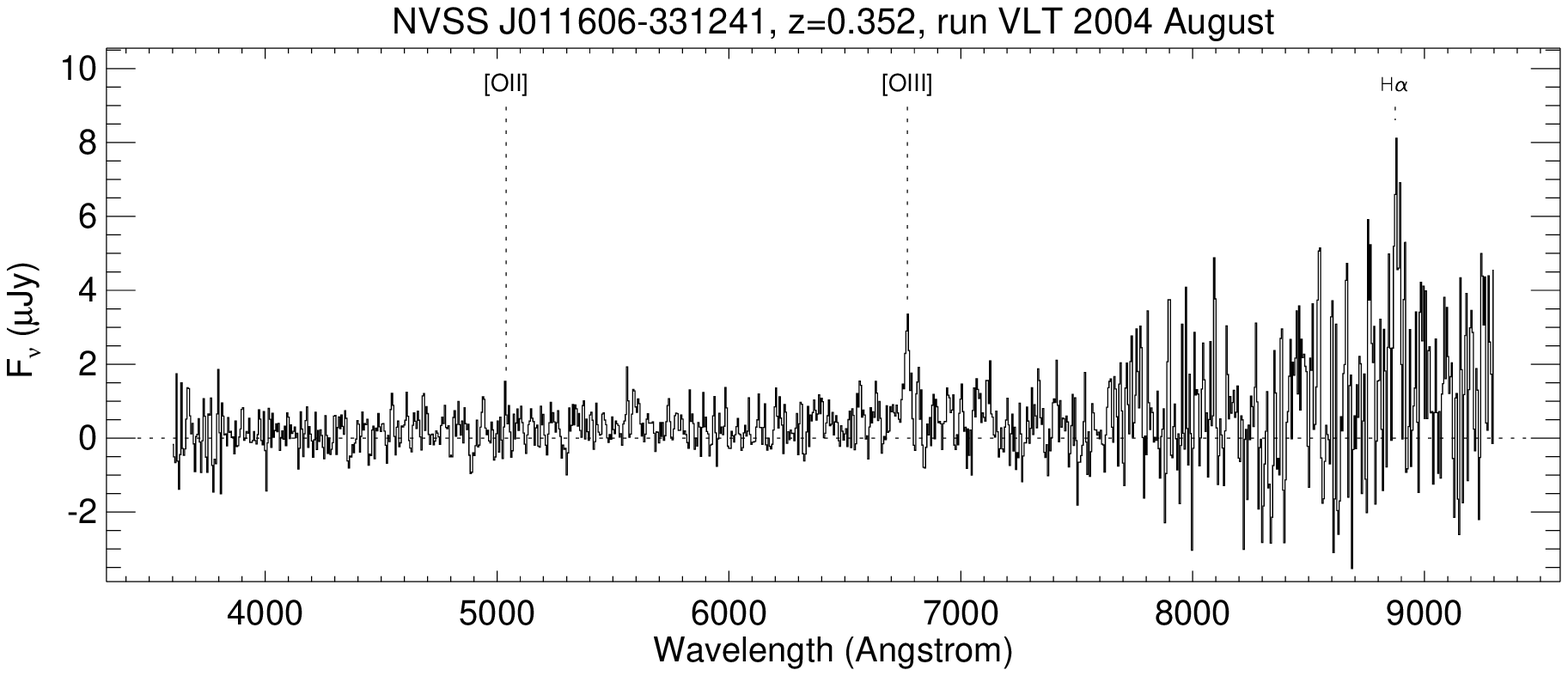,width=8.7cm}
\psfig{file=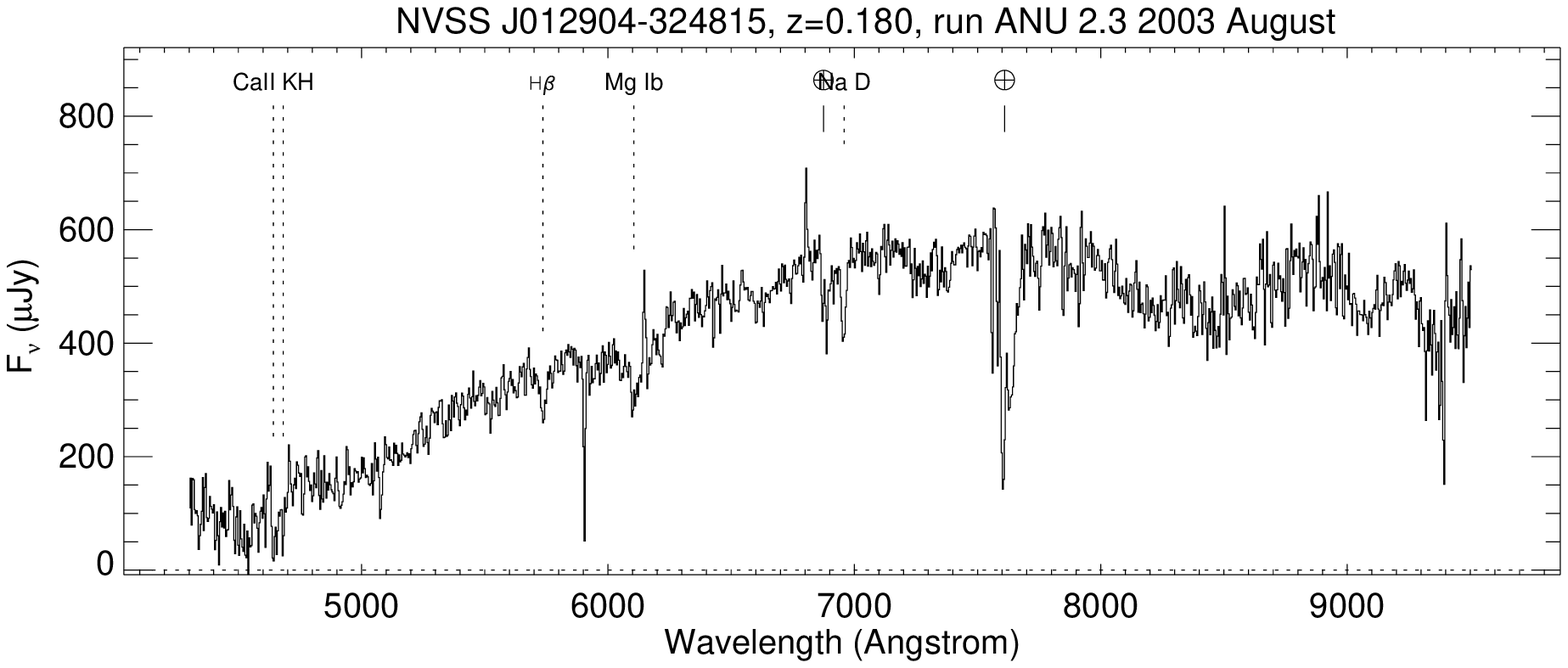,width=8.7cm}
\psfig{file=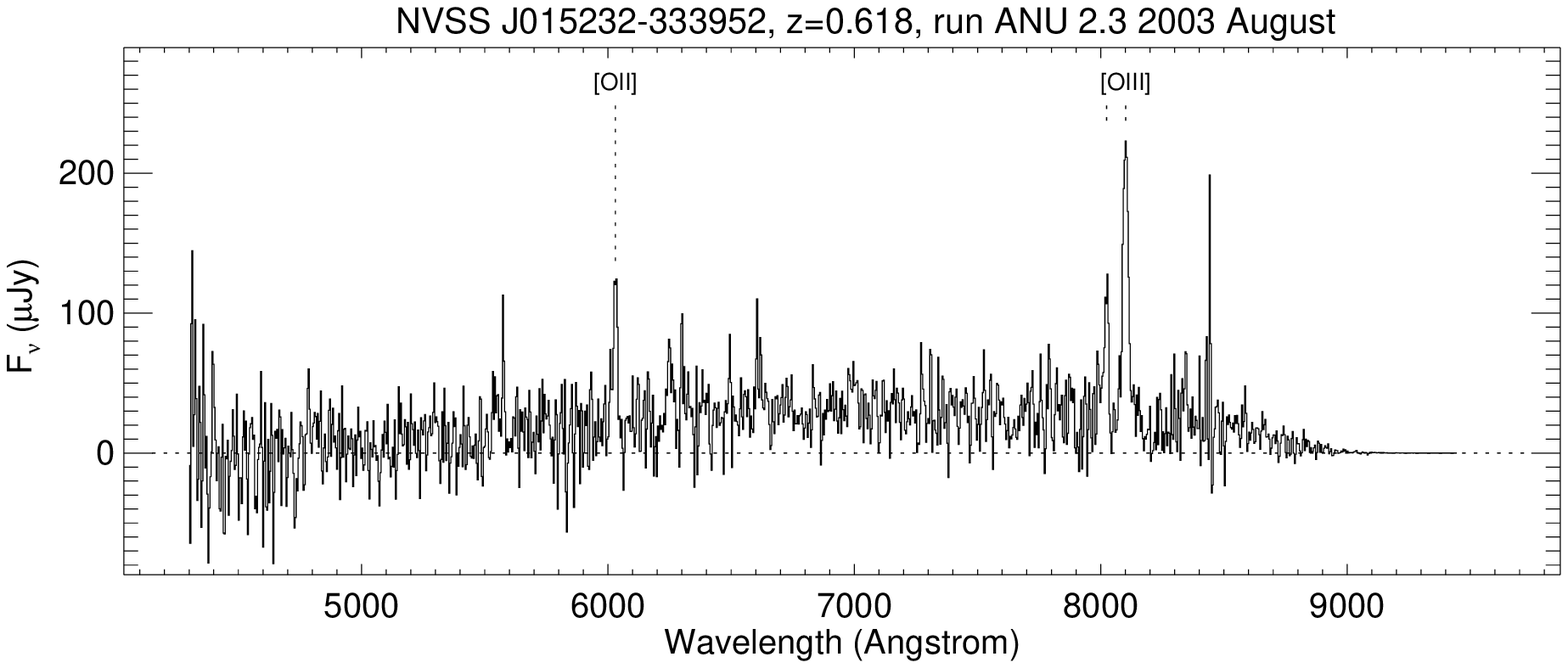,width=8.7cm}
\end{figure} 
\begin{figure}
\psfig{file=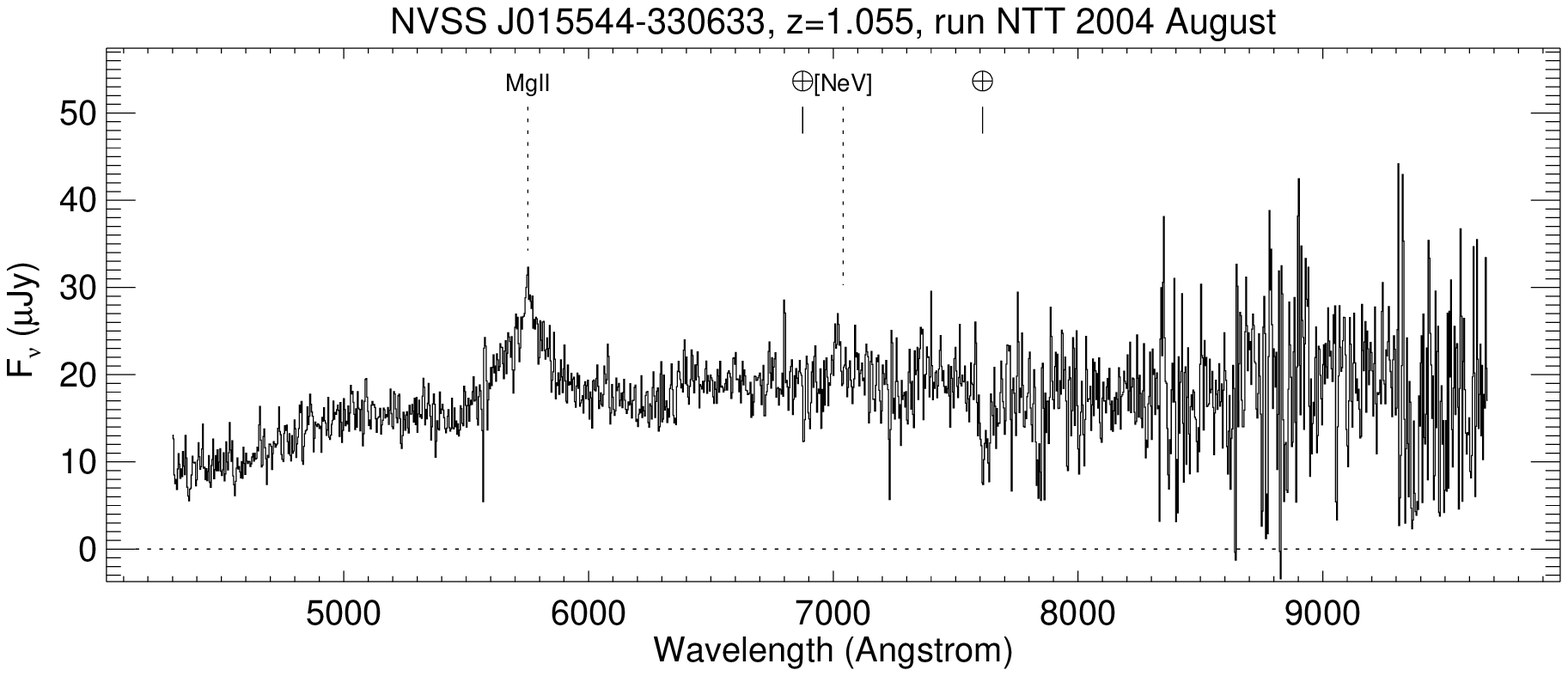,width=8.7cm}
\psfig{file=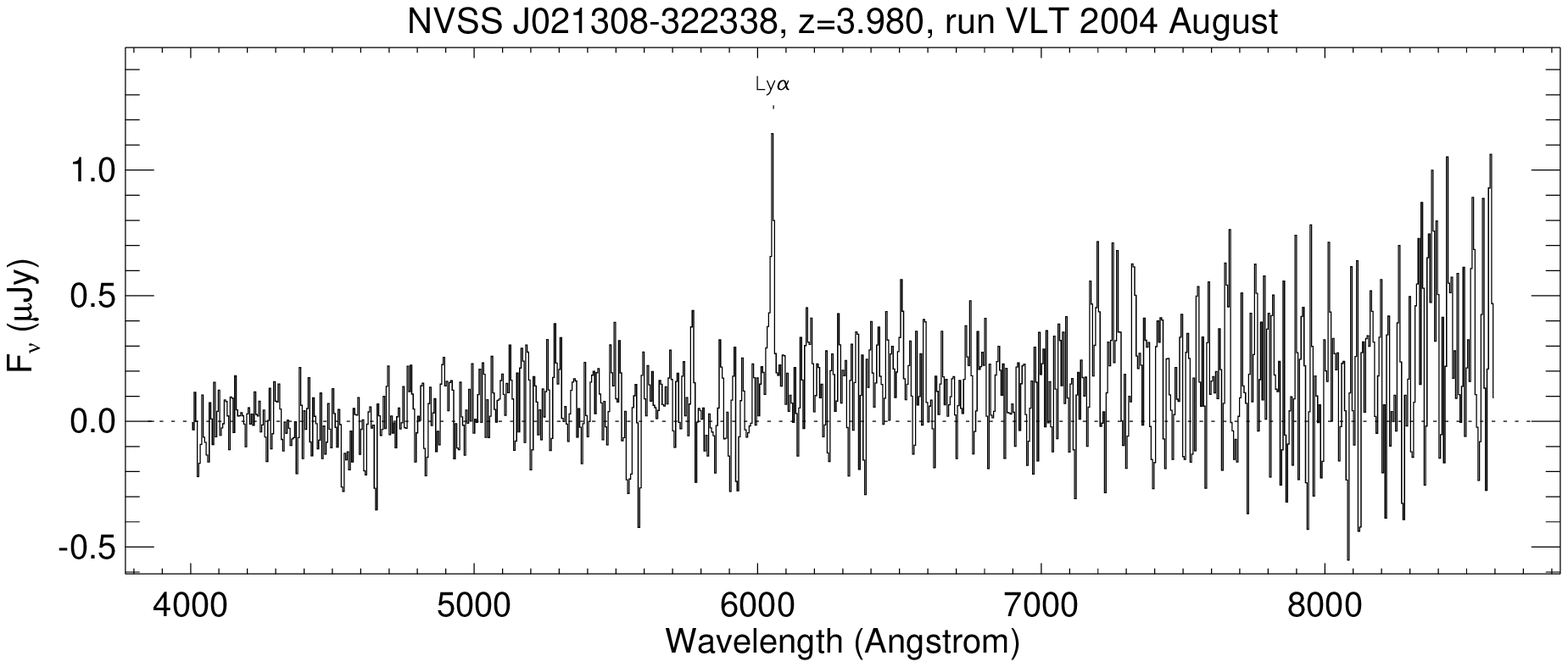,width=8.7cm}
\psfig{file=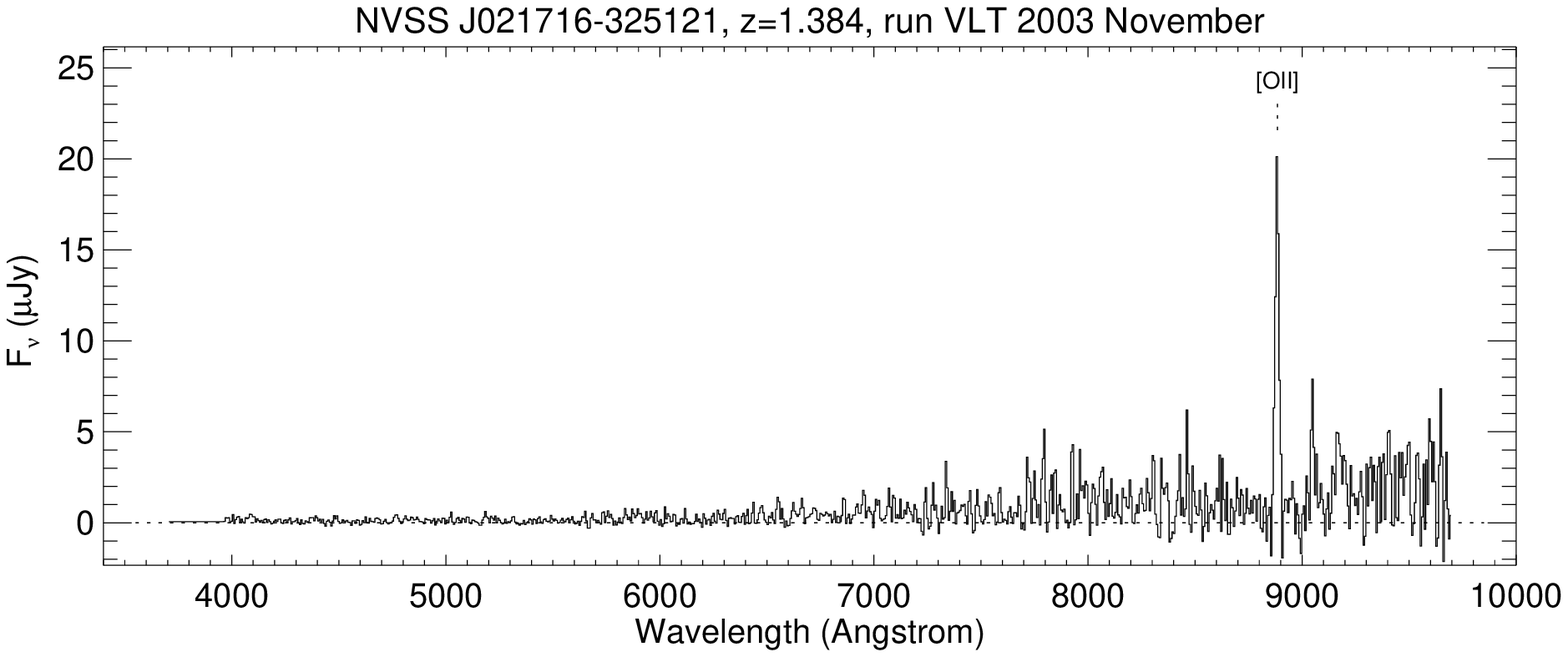,width=8.7cm}
\psfig{file=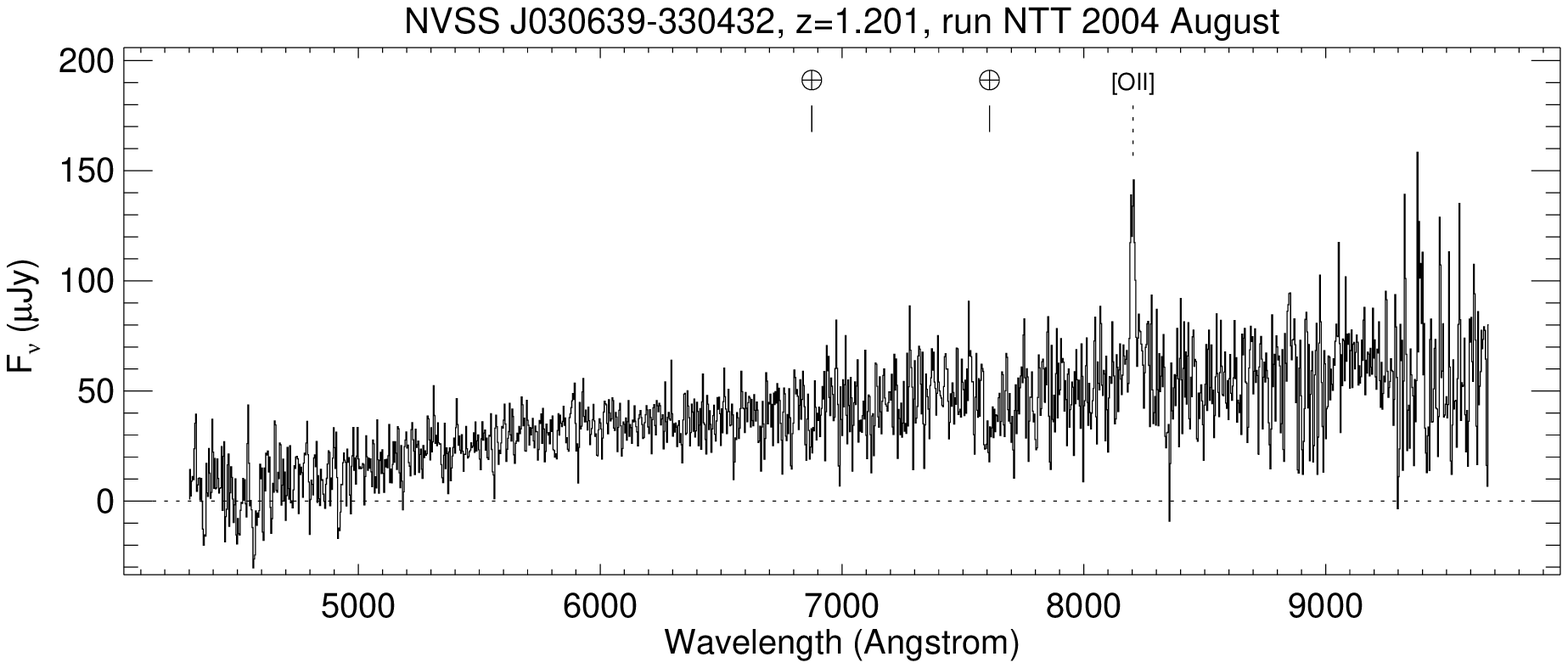,width=8.7cm}
\psfig{file=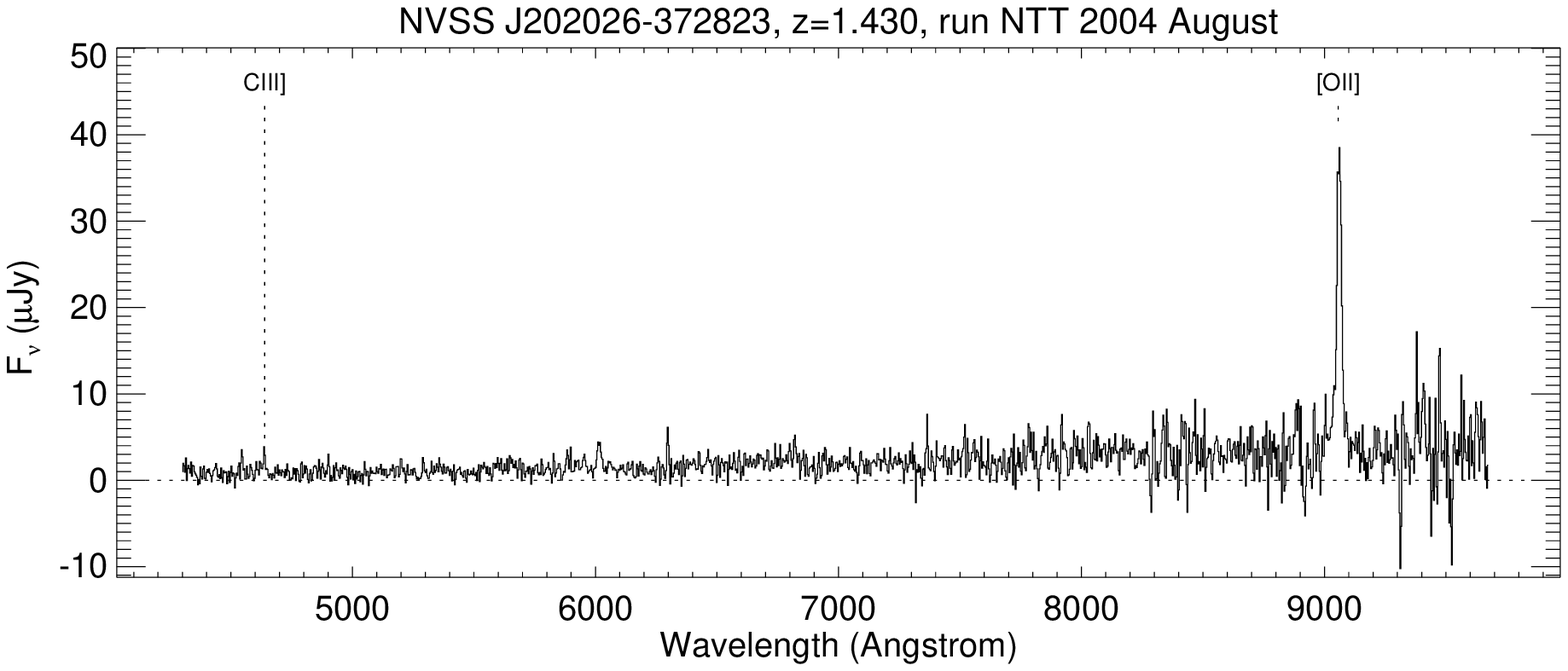,width=8.7cm}
\psfig{file=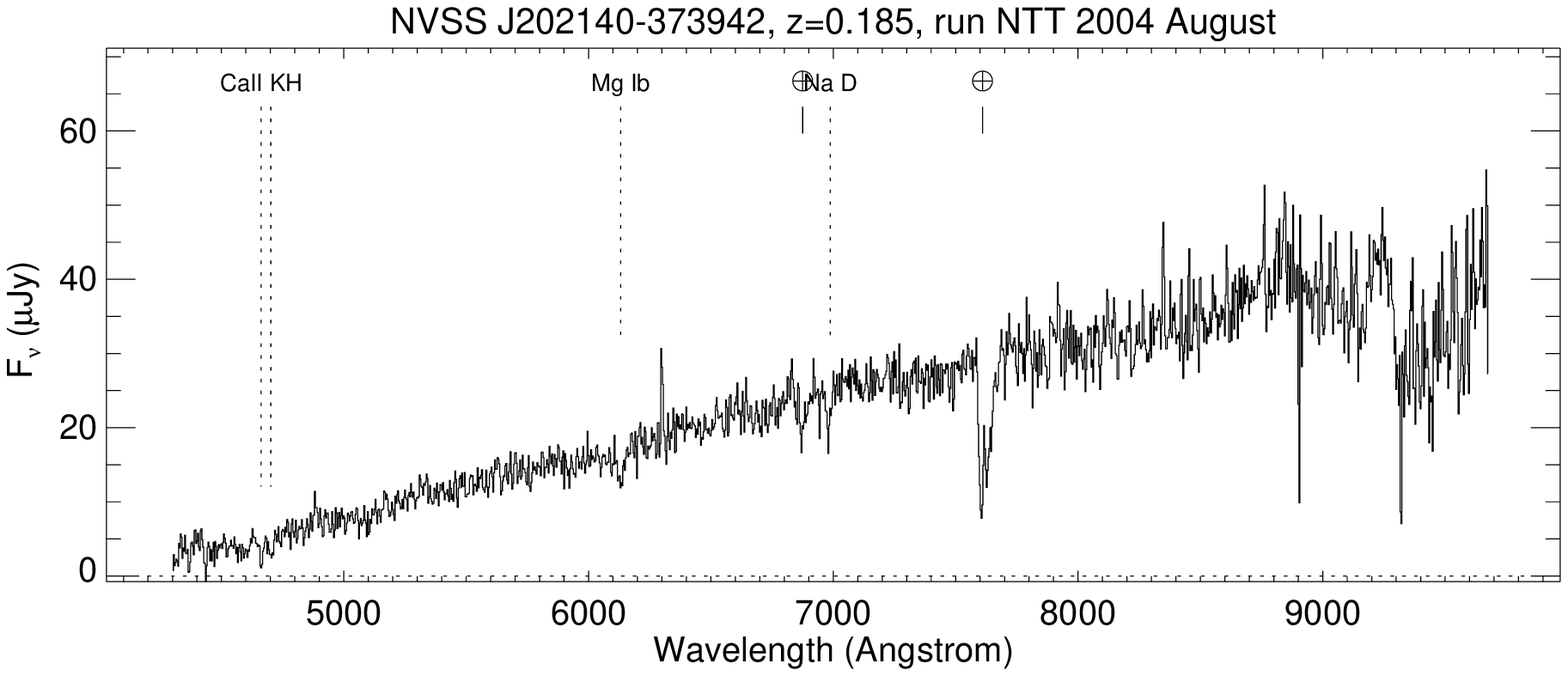,width=8.7cm}
\psfig{file=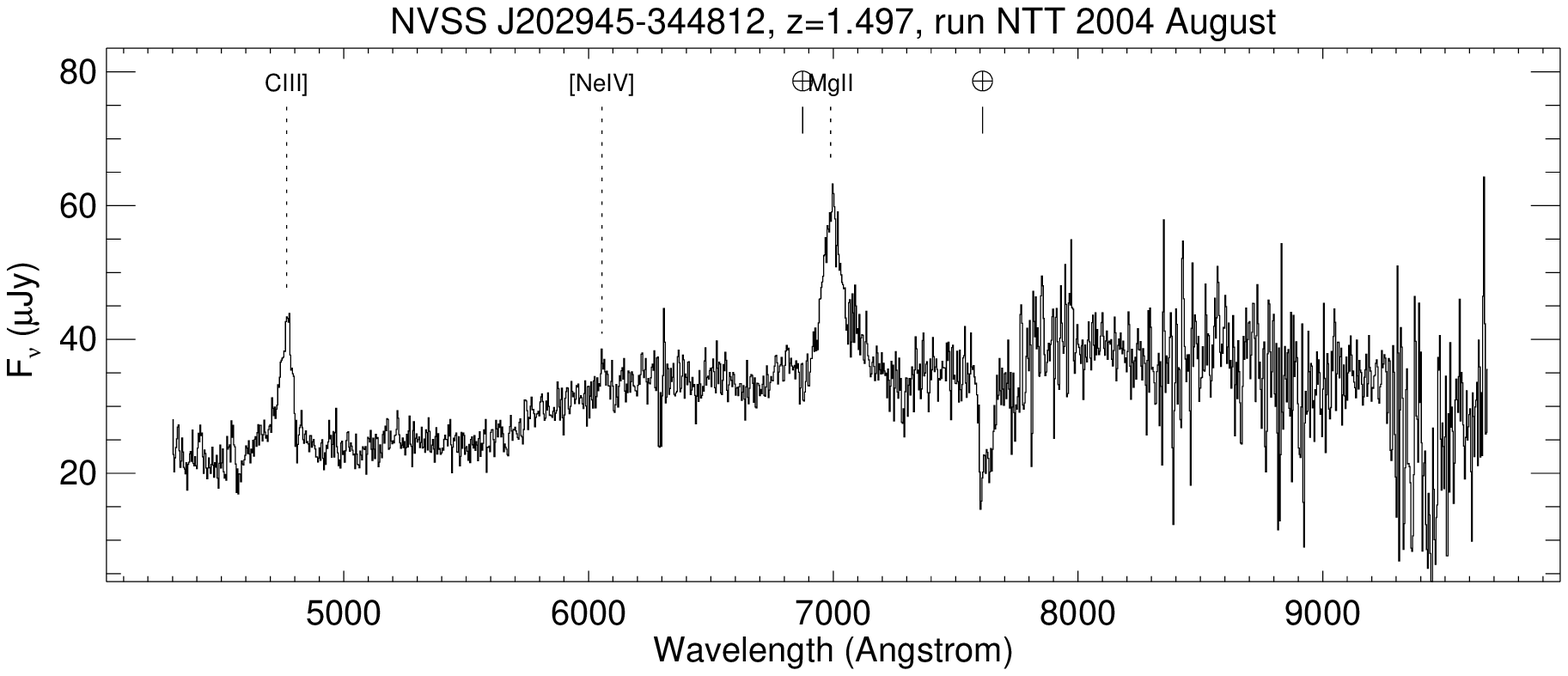,width=8.7cm}
\end{figure} 
\begin{figure}
\psfig{file=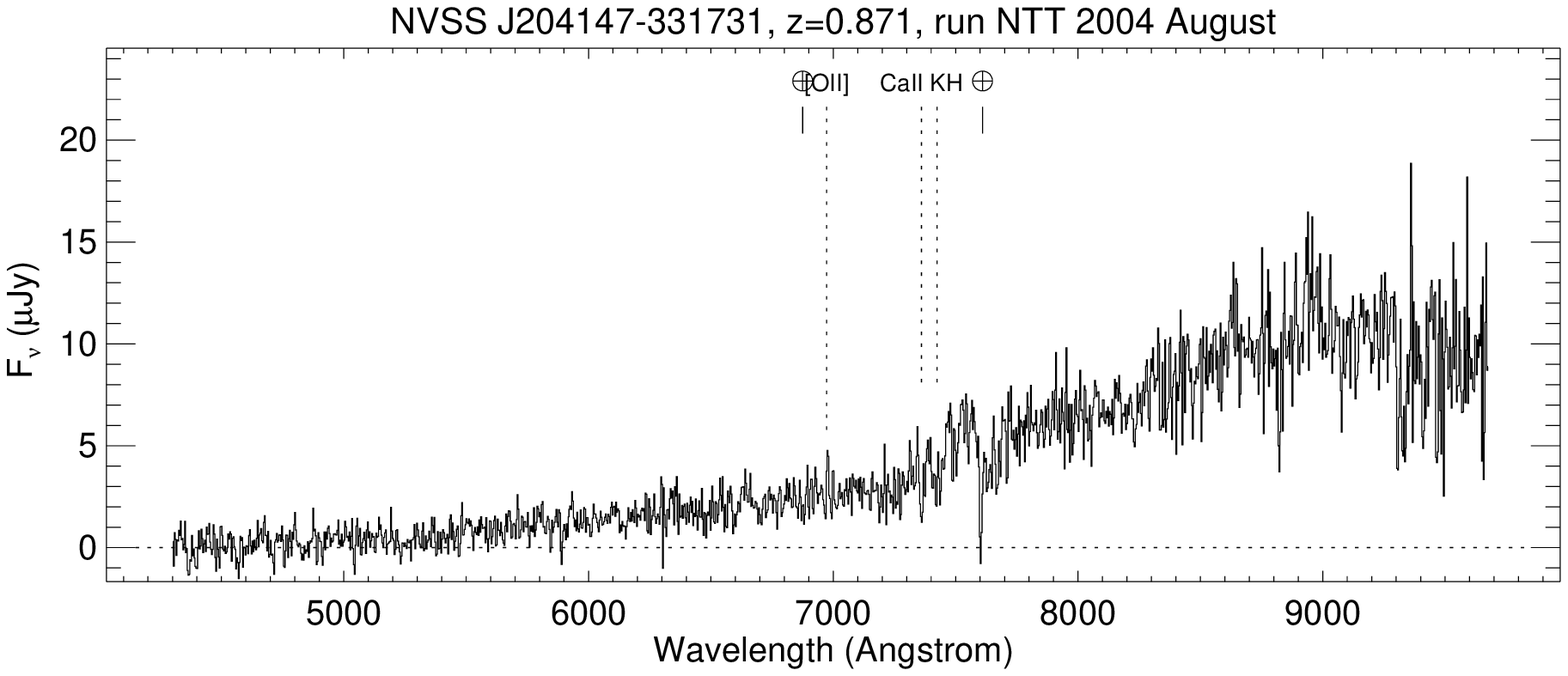,width=8.7cm}
\psfig{file=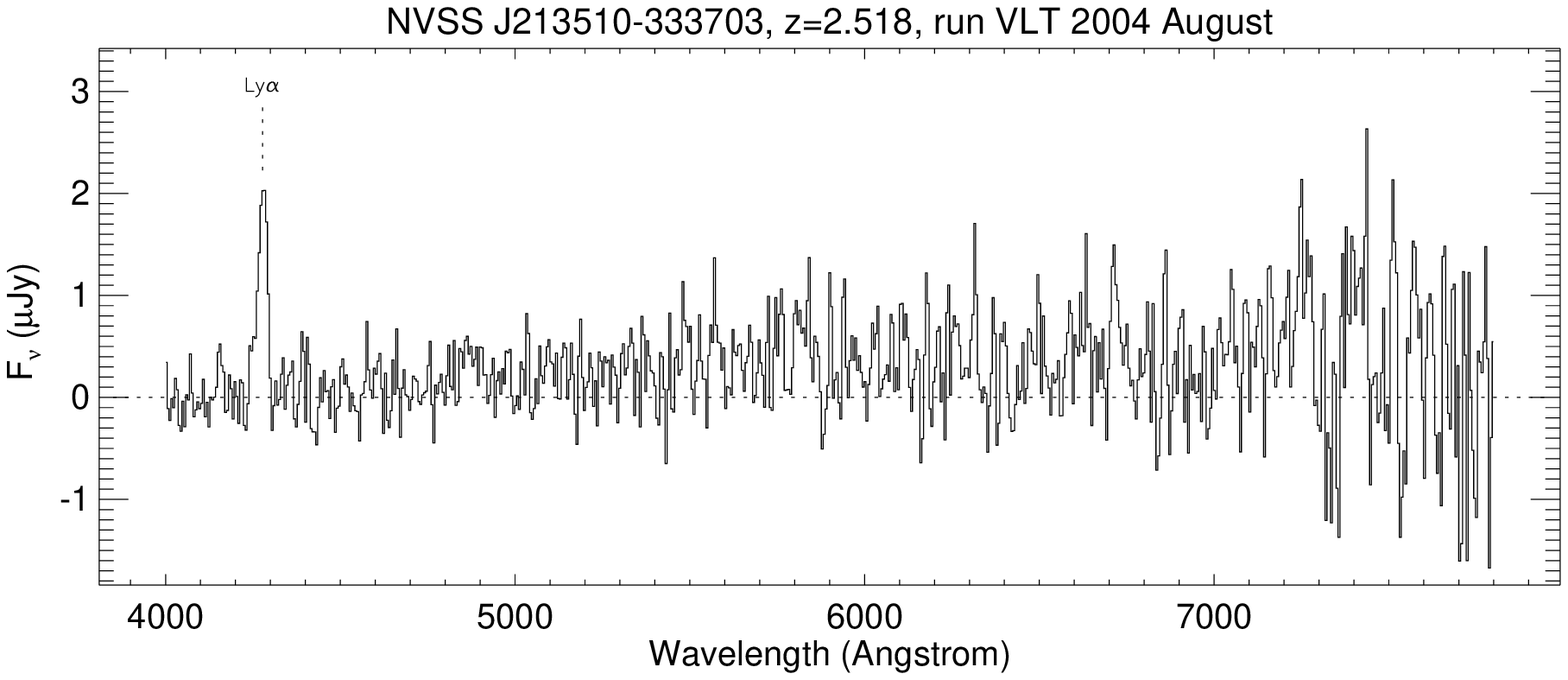,width=8.7cm}
\psfig{file=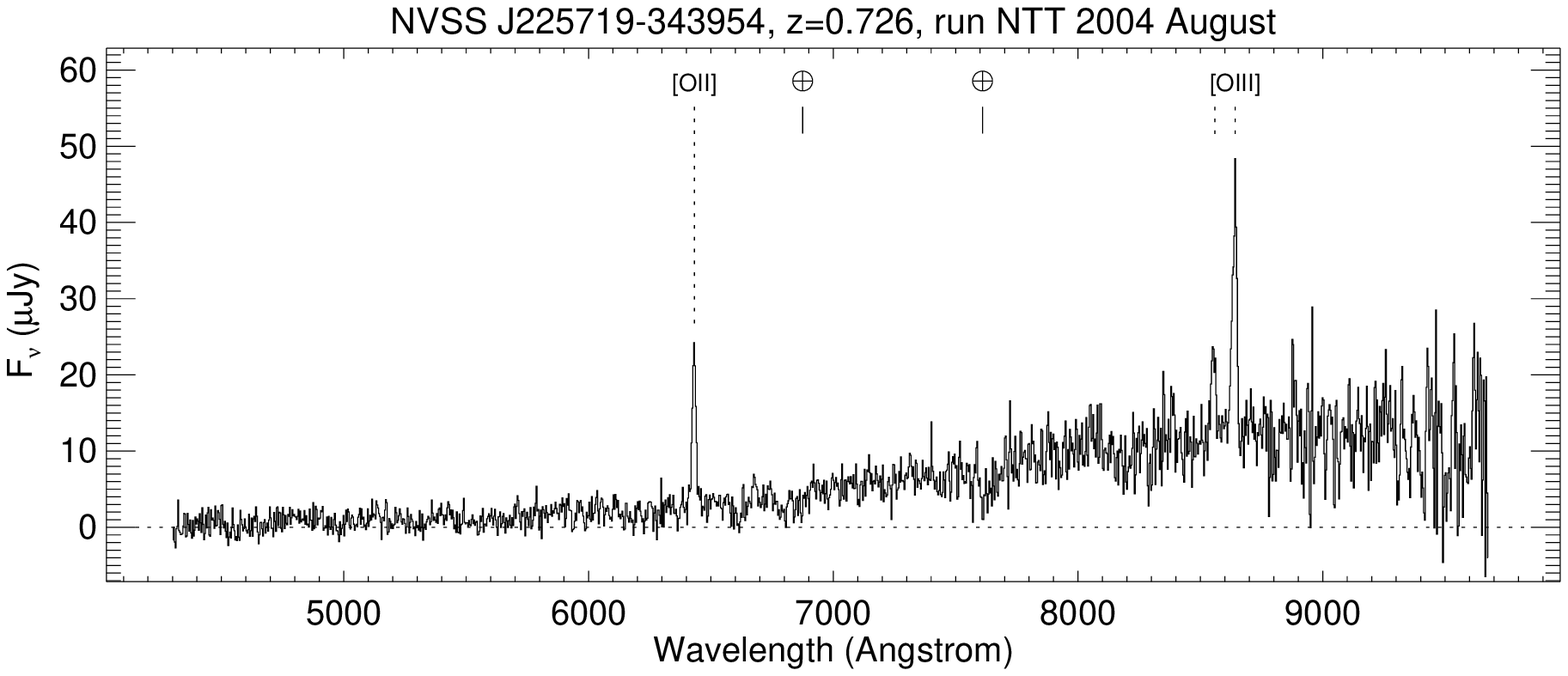,width=8.7cm}
\psfig{file=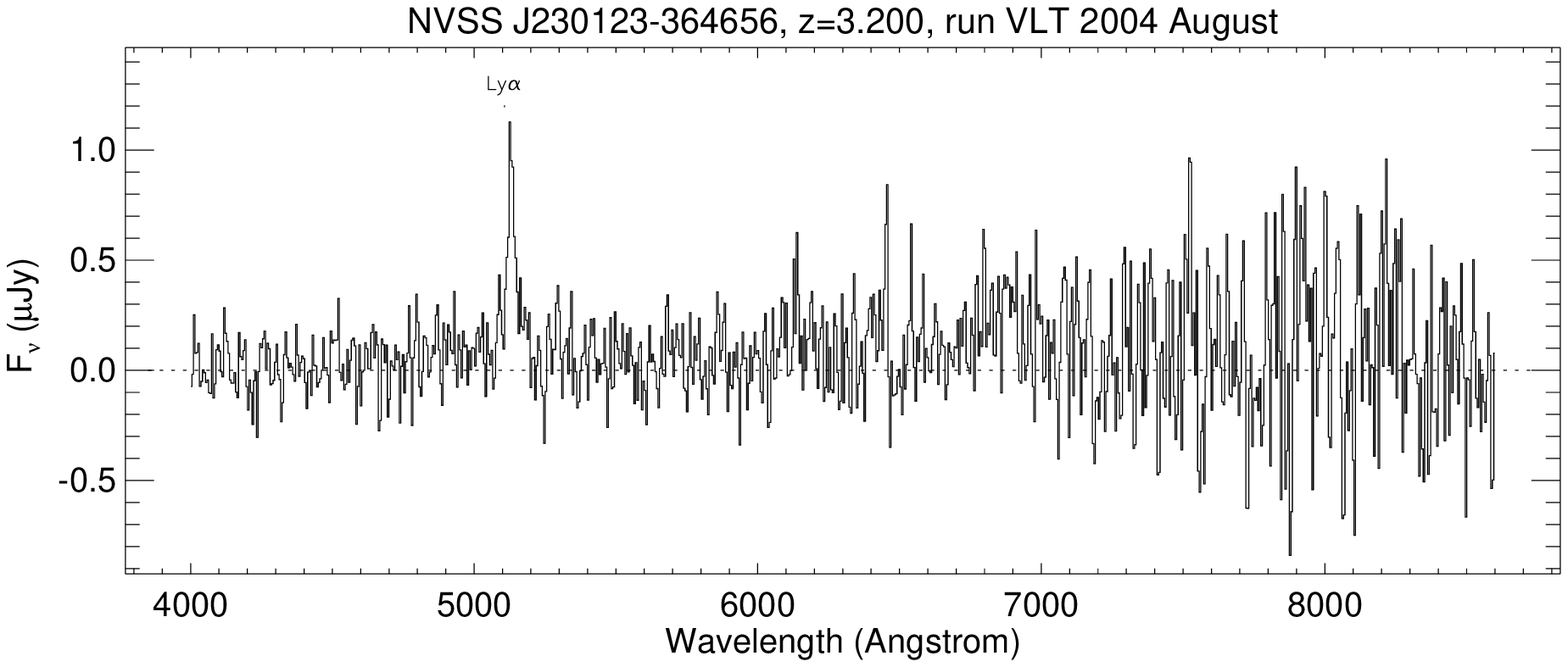,width=8.7cm}
\psfig{file=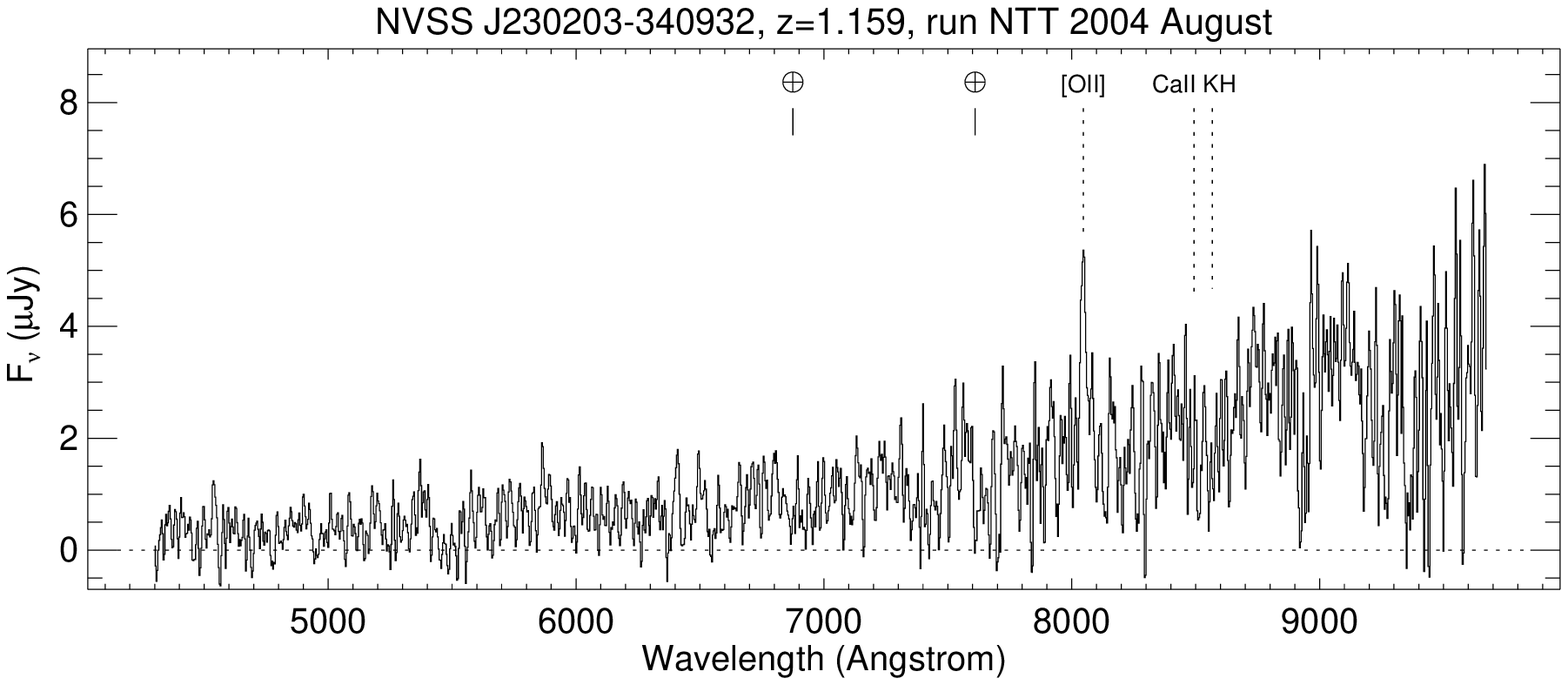,width=8.7cm}
\psfig{file=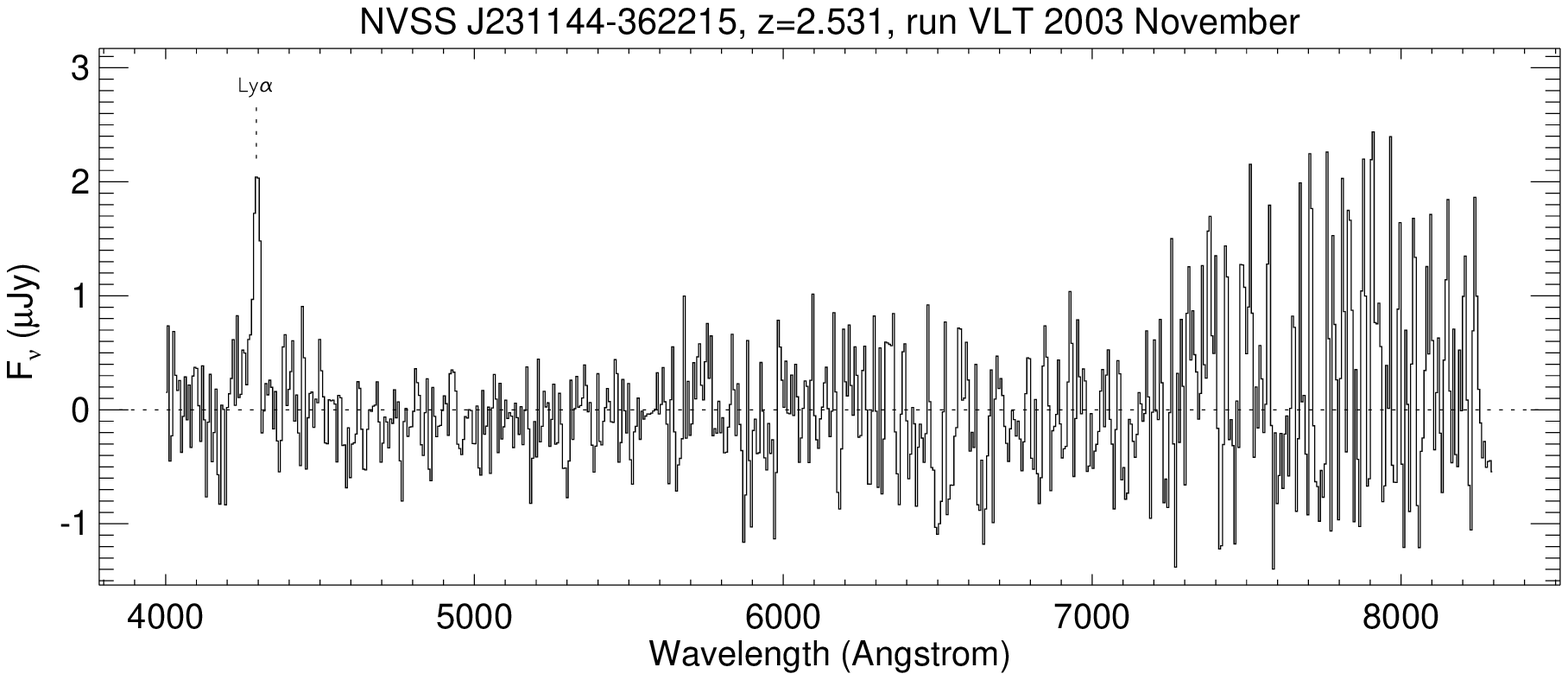,width=8.7cm}
\psfig{file=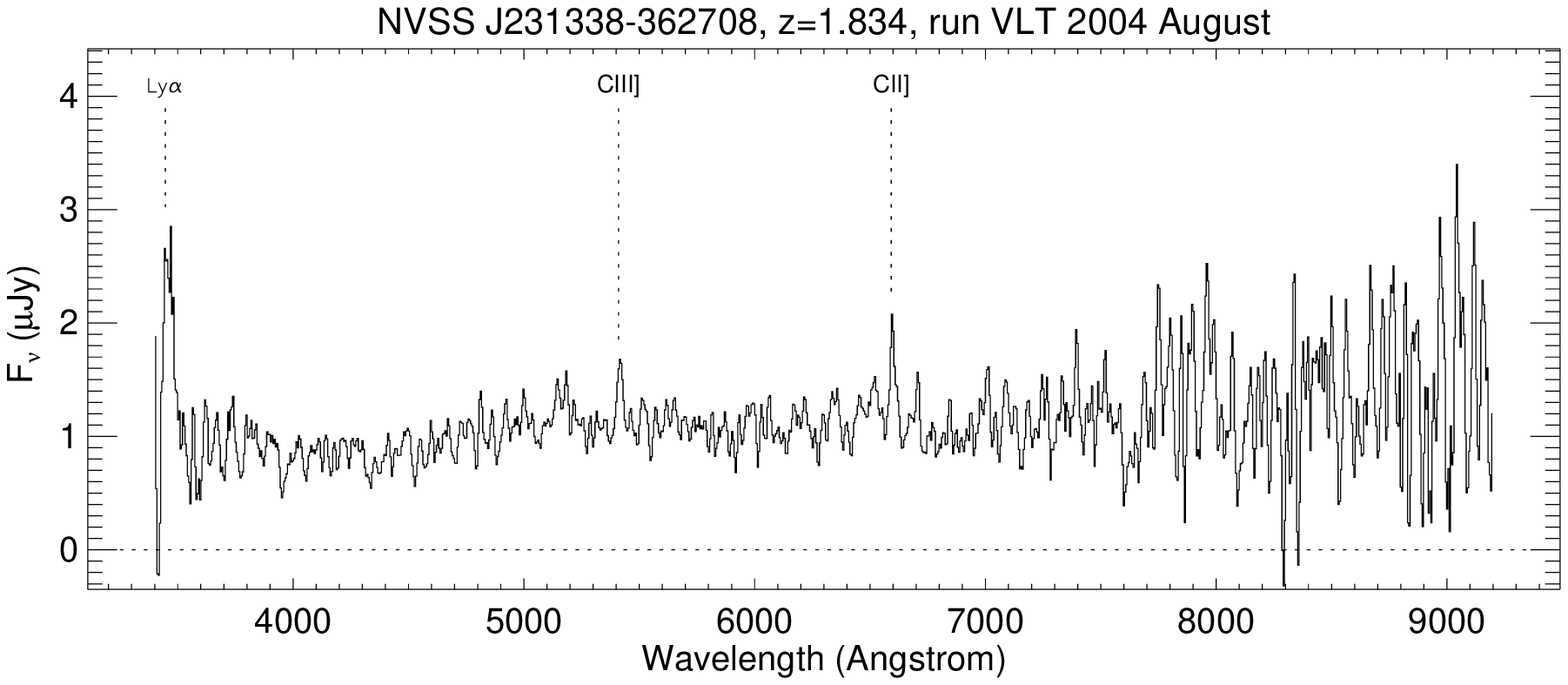,width=8.7cm}
\end{figure} 
\begin{figure}
\psfig{file=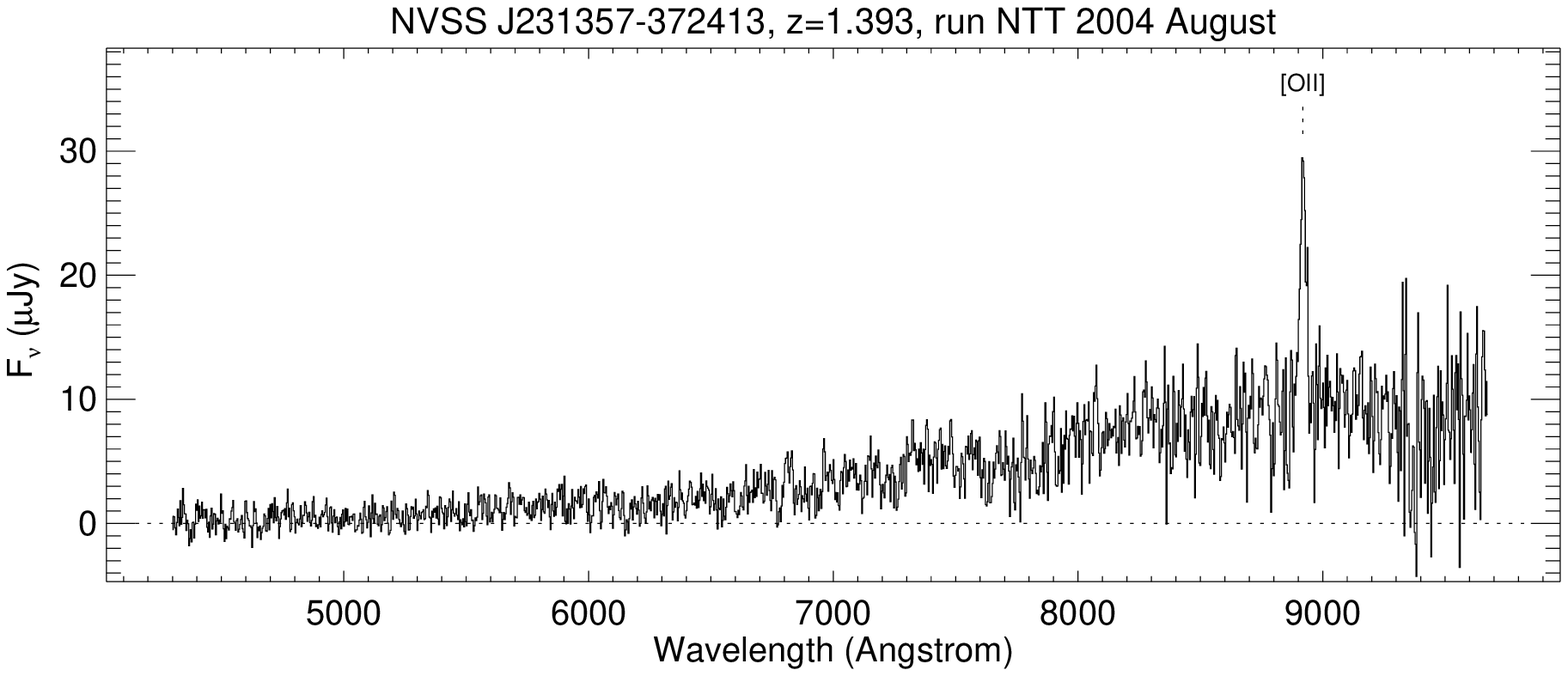,width=8.7cm}
\psfig{file=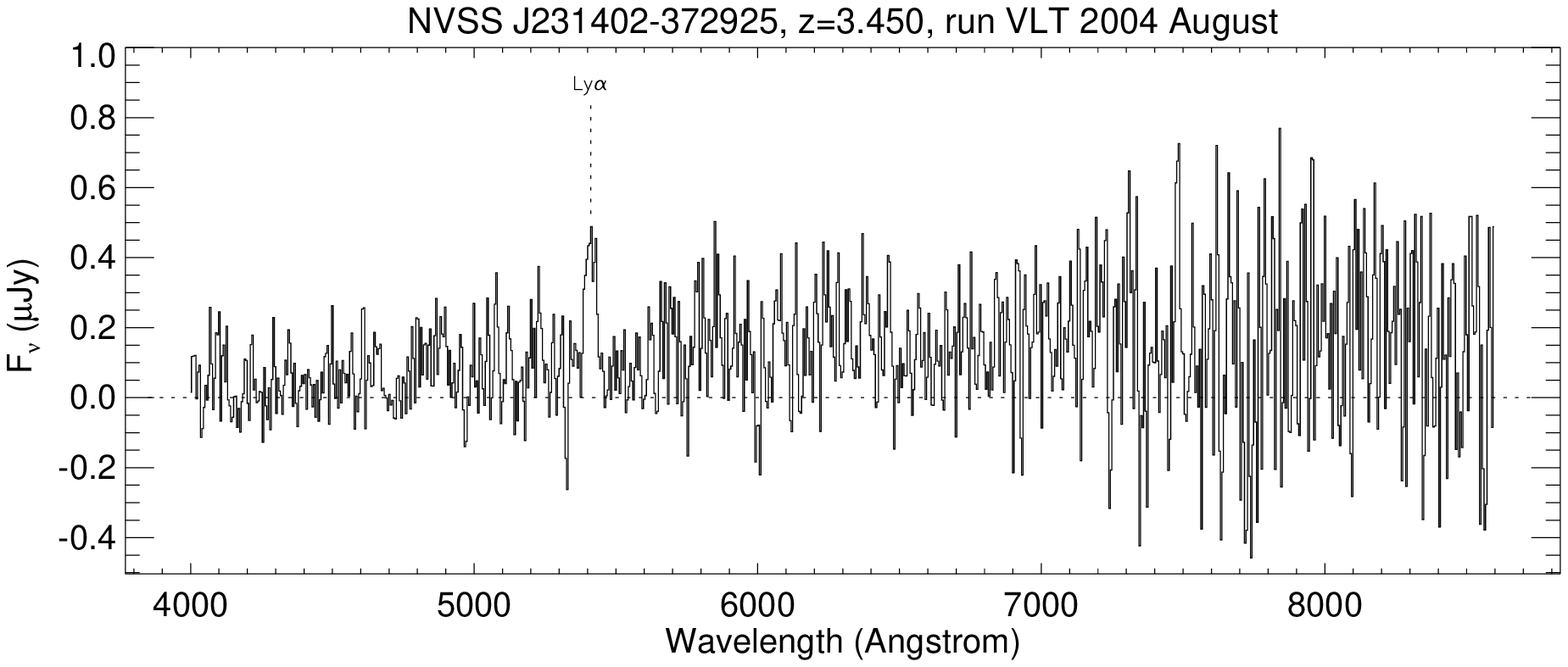,width=8.7cm}
\psfig{file=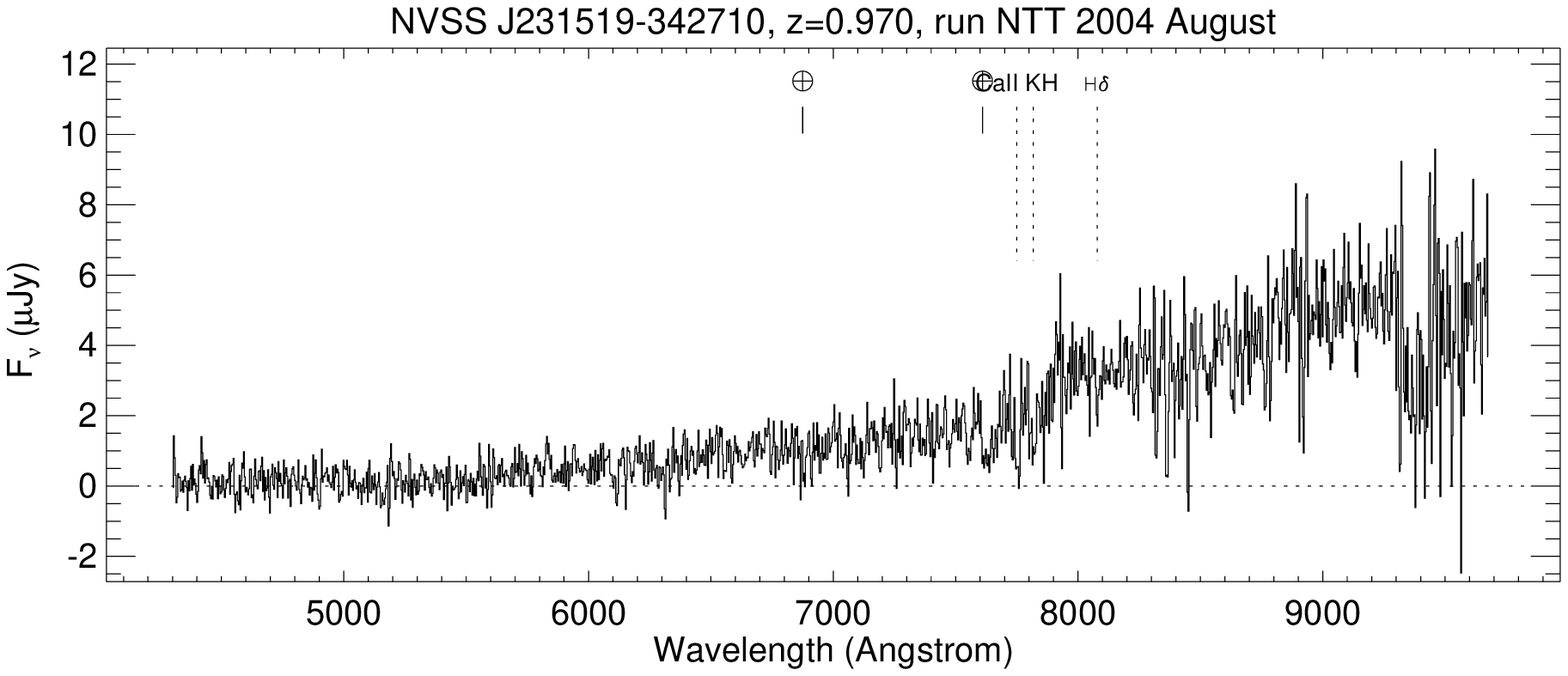,width=8.7cm}
\psfig{file=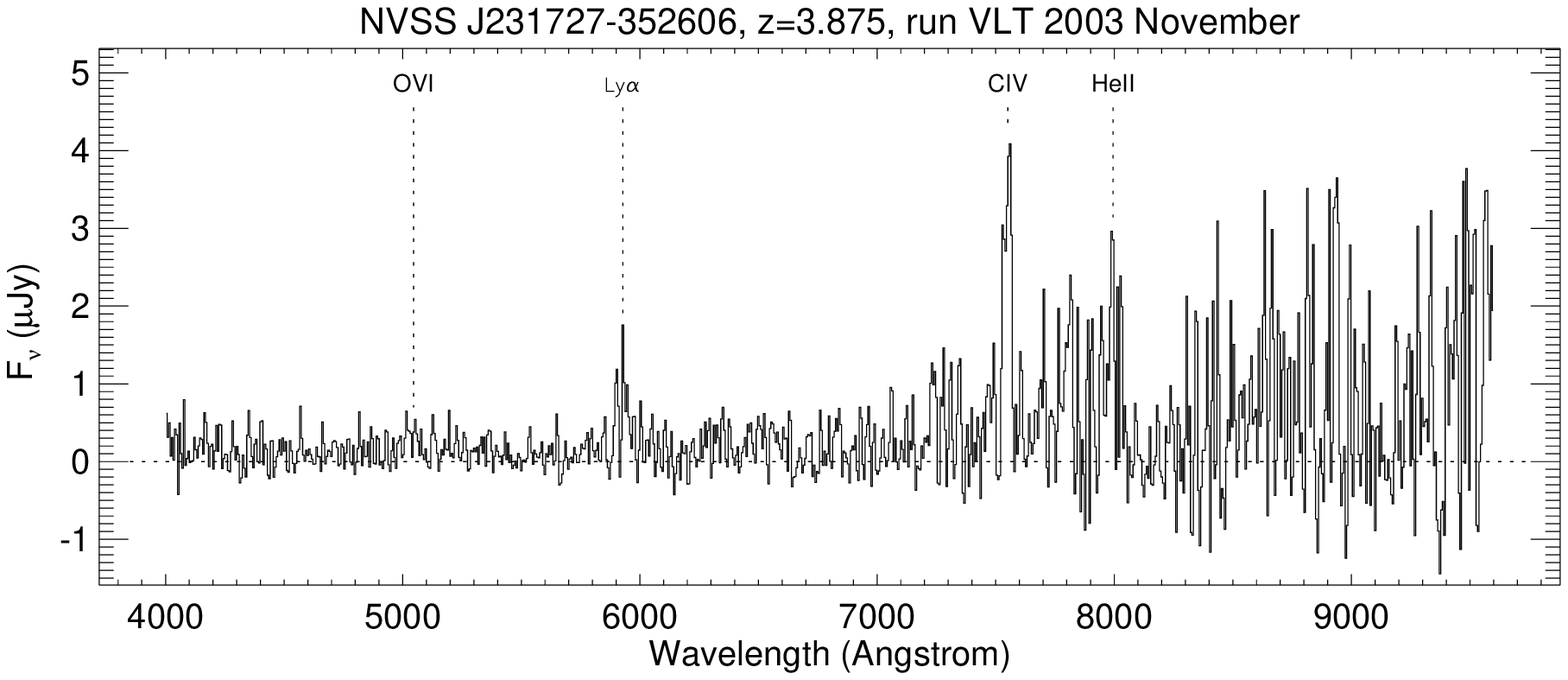,width=8.7cm}
\psfig{file=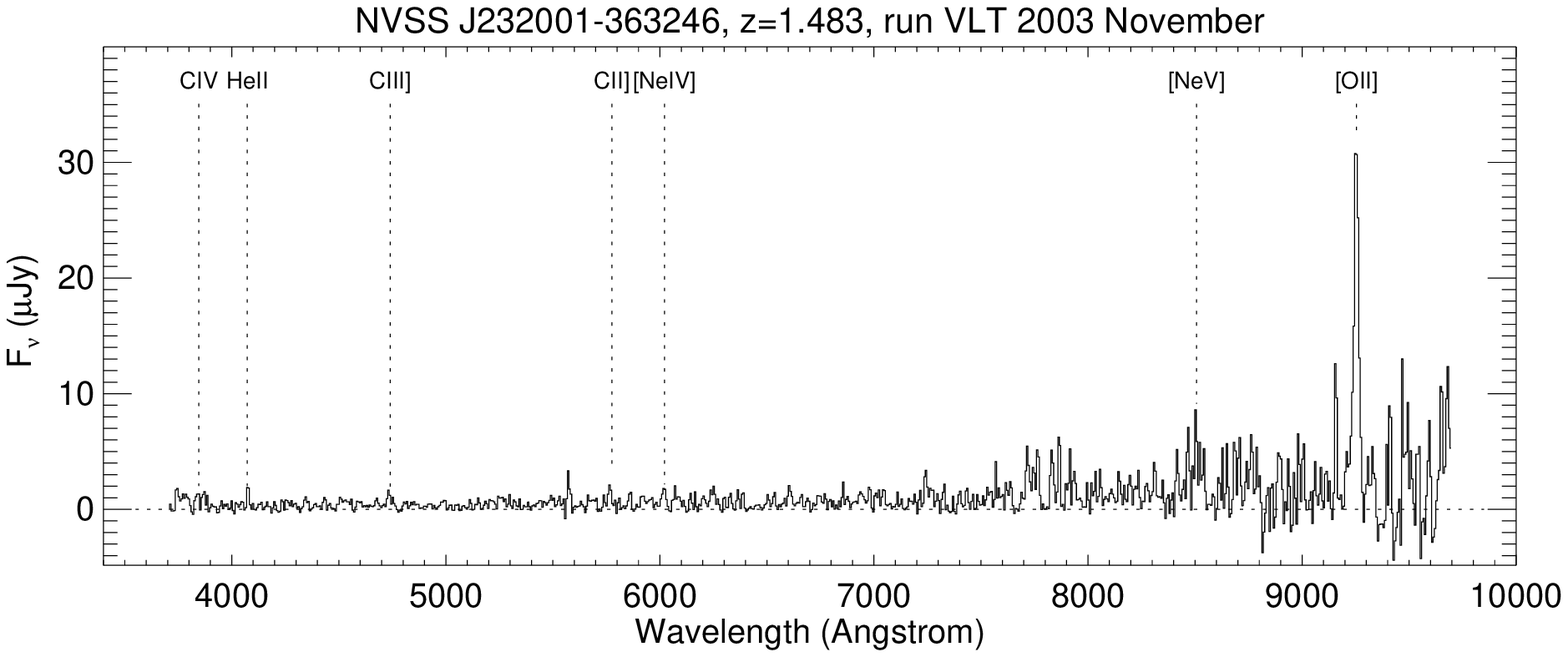,width=8.7cm}
\psfig{file=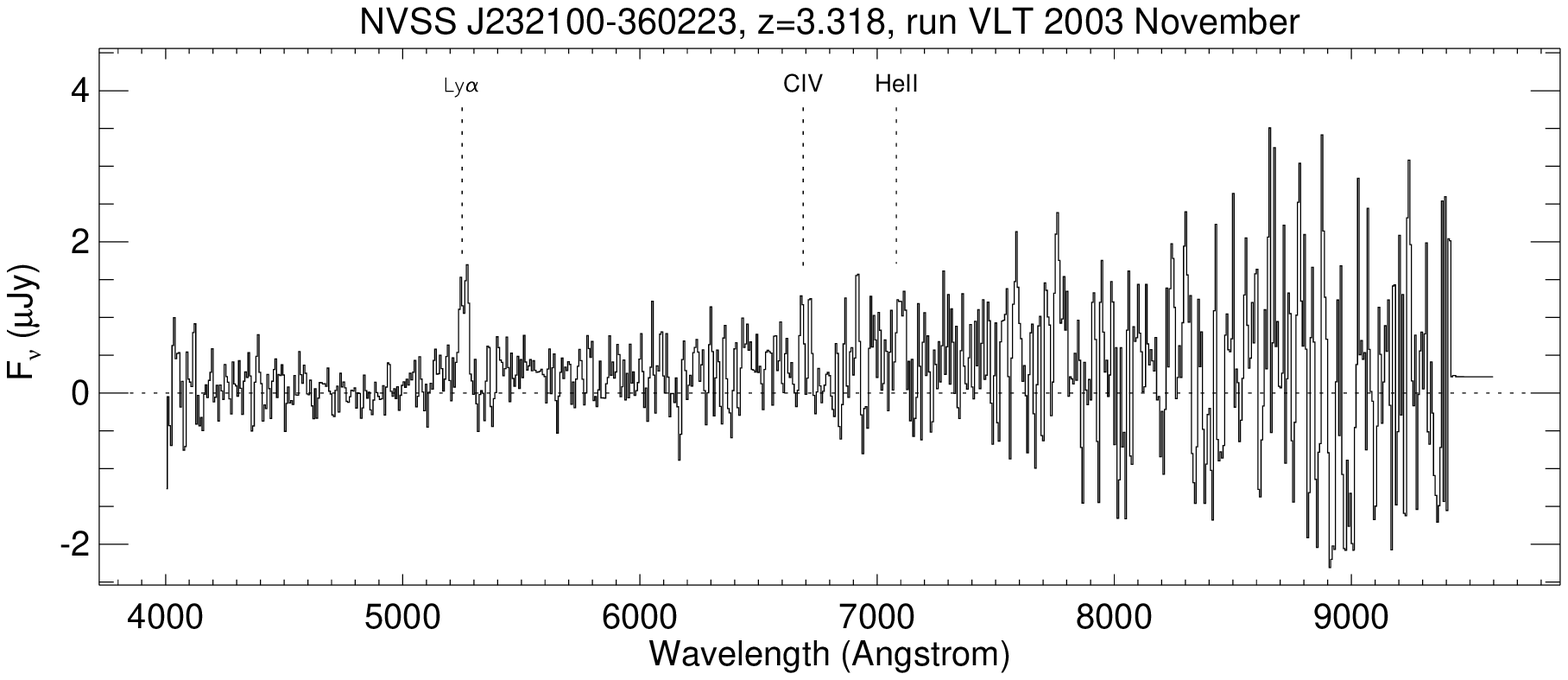,width=8.7cm}
\psfig{file=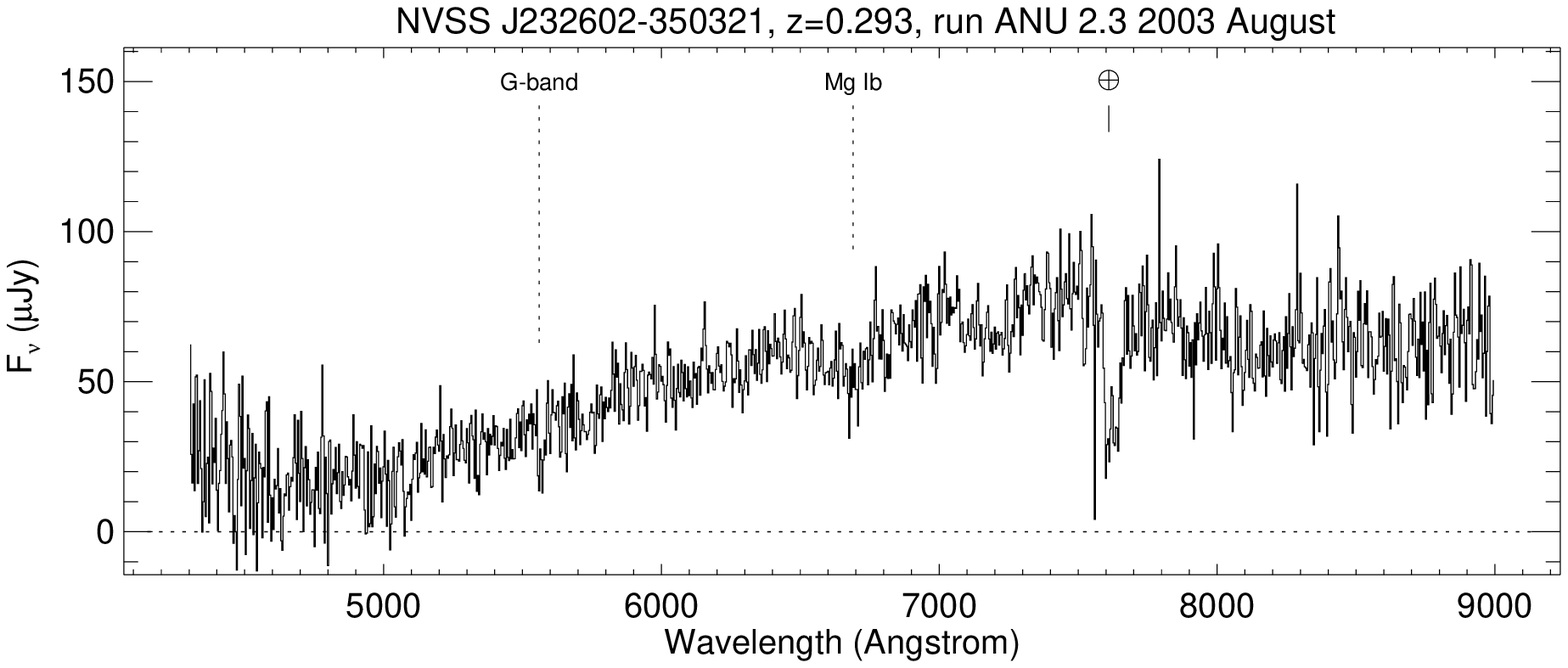,width=8.7cm}
\end{figure} 
\begin{figure}
\psfig{file=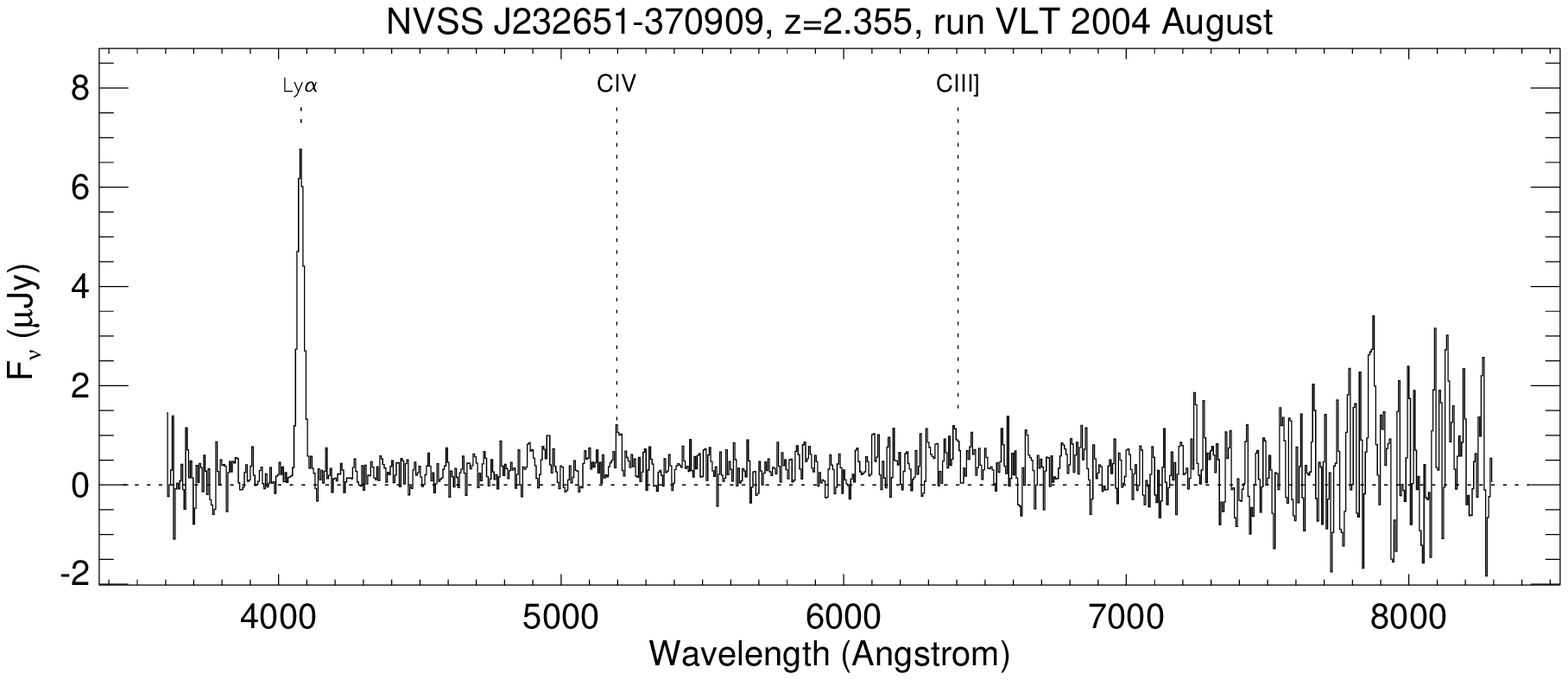,width=8.7cm}
\psfig{file=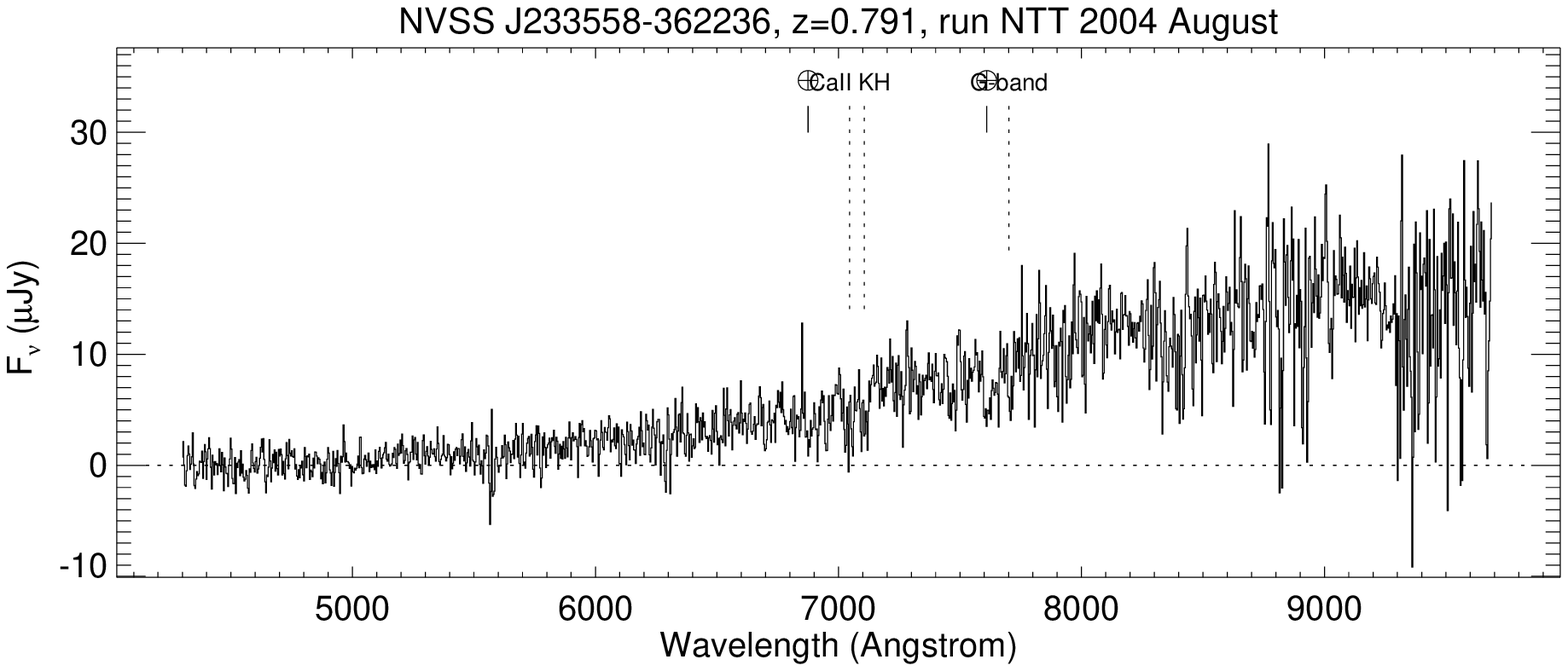,width=8.7cm}
\psfig{file=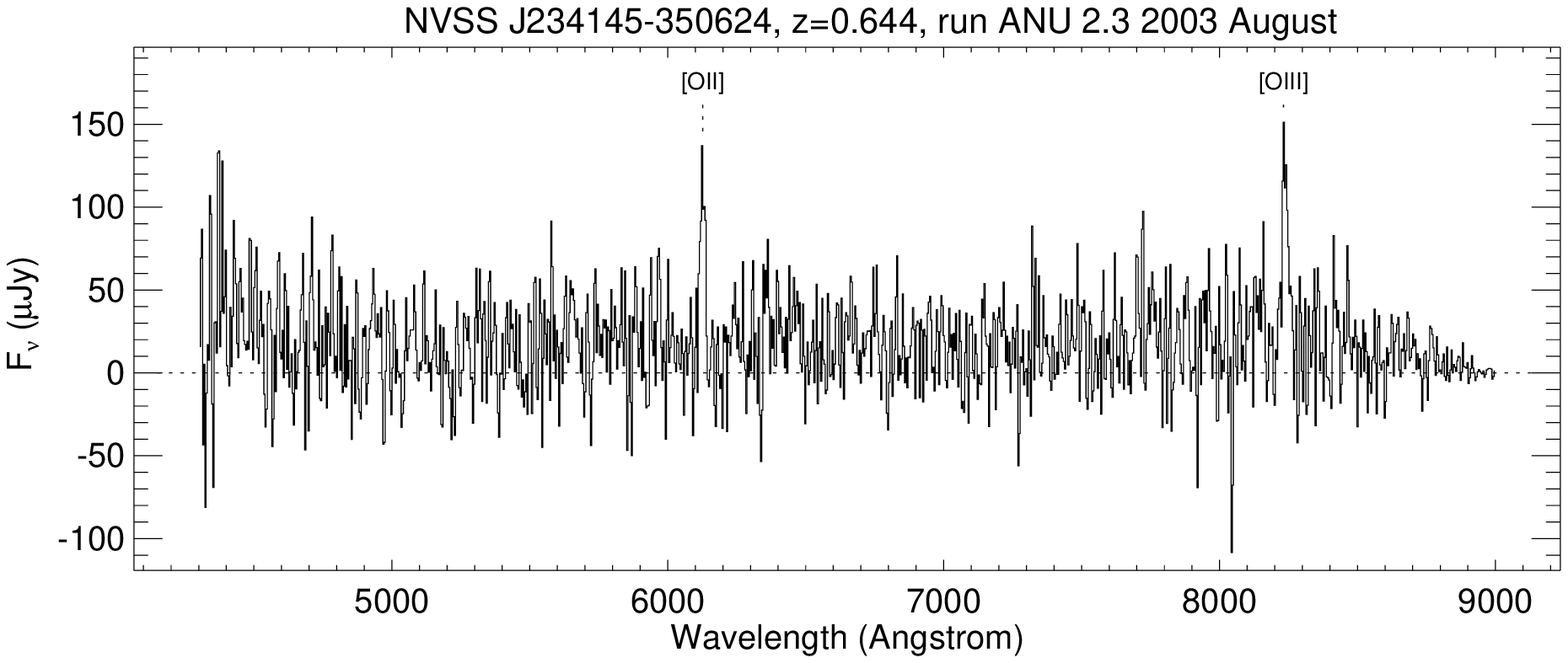,width=8.7cm}
\psfig{file=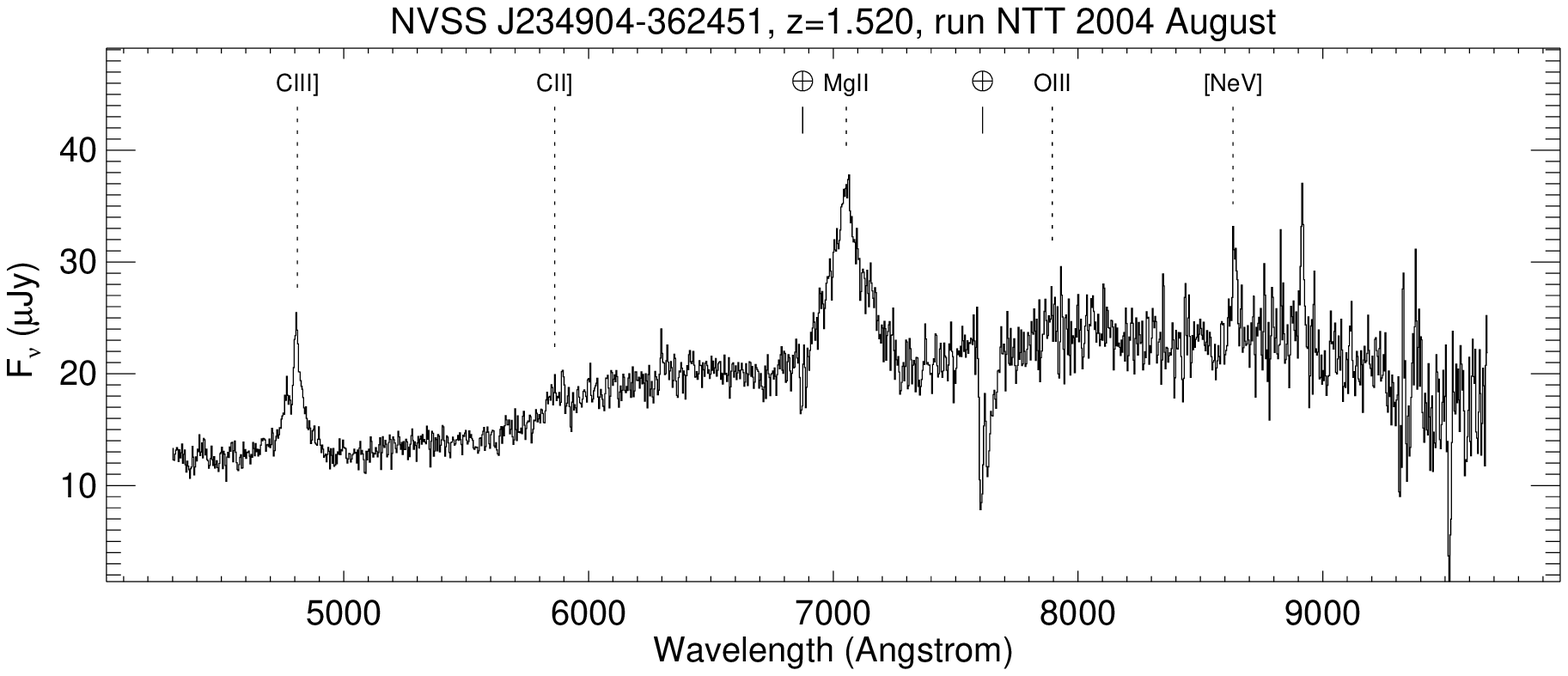,width=8.7cm}

\caption{ Spectra of sources from the SUMSS--NVSS USS sample with
prominent features indicated. The positions of the vertical dotted lines
indicate the predicted observed wavelength of the lines at the redshift
quoted above each individual spectrum, and not the wavelength of the
fitted peak. Differences between the feature and the dotted line thus show
velocity shifts of the lines. The source name, redshift, and observing run
are shown on top of each spectrum.  Telluric OH absorption is indicated by
$\oplus$.
}
\label{1Dspectra}
\end{figure} 

\noindent {\bf NVSS~J002001$-$333408:} The continuum is well detected
(Fig.~\ref{continua}), but we see no emission or absorptions lines. The
rise in the continuum around $\sim$8200\,\AA\ is probably due to the
4000\,\AA\ break at $z\sim 1$.

\noindent {\bf NVSS~J002427$-$325135:} The continuum is well detected
(Fig.~\ref{continua}), but we see no emission or absorption lines. The
rise in the continuum around $\sim$8600\,\AA\ is probably due to the
4000\,\AA\ break at $z\sim 1.2$.

\noindent {\bf NVSS~J021308$-$322338:} At $z=3.976$, this is the
most distant radio galaxy discovered to date from the SUMSS--NVSS
USS sample. The redshift is based on a single emission line at
$\lambda_{\rm obs}=6051$\,\AA. The continuum discontinuity across
the line, and the absence of other lines in our wide spectral
coverage identifies this line as \Lya.

\noindent {\bf NVSS~J021716$-$325121:} The redshift is based on a single emission line, which we identify as \OII, based on the absence of confirming lines if the line were \Lya, \OIII\ or \Ha, and the presence of clear underlying continuum emission.

\noindent {\bf NVSS~J030639$-$330432:} The redshift is based on a single emission line, which we identify as \OII, based on the absence of confirming lines if the line were \Lya, \OIII\ or \Ha, and the presence of clear underlying continuum emission.

\noindent {\bf NVSS~J202518$-$355834:} We observed this source at two
different position angles to ensure we covered all possible
identifications. However, the central object coincident with the radio
source remains undetected in our VLT spectra.

\noindent {\bf NVSS~J204420$-$334948:} No line or continuum emission was
detected in the 3600s FORS2 spectrum. Because of its extremely steep radio
spectrum ($\alpha_{843}^{1400}=-1.60$) and favourable RA at the time
of the FORS1 observations, we obtained a further 8100s spectrum.
The source remains undetected, indicating either that redshift
determinations of such very faint sources are beyond the capabilities of
present-day optical spectrographs, or that this source may be at
$z\simgt$7.

\noindent {\bf NVSS~J213510$-$333703:} The redshift is based on a single emission line, which we interpret as \Lya, based on the absence of confirming lines if the line were \OII, \OIII\ or \Ha, and the faint continuum discontinuity across the line.

\noindent {\bf NVSS~J230123$-$364656:} The redshift is based on a single emission line, which we interpret as \Lya, based on the absence of confirming lines if the line were \OII, \OIII\ or \Ha, and the faint continuum discontinuity across the line.

\noindent {\bf NVSS~J231144$-$362215:} The redshift is based on a single emission line, which we interpret as \Lya, based on the absence of confirming lines if the line were \OII, \OIII\ or \Ha, and absence of continuum emission.

\noindent {\bf NVSS~J231338$-$362708:} This is a typical example of a
source in the `redshift desert', with \Lya\ detected at the very edge of
the spectral coverage. The \Lya\ line is diffuse and extended, and the
continuum emission is very strong. The redshift is confirmed by carbon
lines.

\noindent {\bf NVSS~J231341$-$372504:} The continuum is well detected
(Fig.~\ref{continua}), but we see no emission or absorption lines. The
rise in the continuum around $\sim$7500\,\AA\ is probably due to the
4000\,\AA\ break at $z\sim 0.9$.

\noindent {\bf NVSS~J231357$-$372413:} The redshift is based on a single emission line, which we identify as \OII, based on the absence of confirming lines if the line were \Lya, \OIII\ or \Ha, and the presence of clear underlying continuum emission.

\noindent {\bf NVSS~J231402$-$372925} The redshift is based on a single emission line, which we interpret as \Lya, based on the absence of confirming lines if the line were \OII, \OIII\ or \Ha, and the continuum discontinuity across the line.

\noindent {\bf NVSS~J232001$-$363246:} This source is at lower redshift
than expected from the $K-z$ relation (see also \S 5.2.3). Both the
redshift and identification of the radio source are secure. 

\noindent {\bf NVSS~J232219$-$355816:} Because no optical counterpart was
detected down to $I\sim25$, we did not attempt to obtain a spectrum.

\noindent {\bf NVSS~J235137$-$362632:} Because no optical counterpart was
detected down to $I\sim25$, we did not attempt to obtain a spectrum.

\section{Efficiency of the host galaxy identification procedure} In
paper~I, we have used 5--10\arcsec\ resolution radio images to
identify the host galaxies of the USS sources in deep $K-$band
images (reaching $K=21$ for the faintest sources). Previous USS
searches have mostly used $\sim$1\arcsec\ resolution radio maps
\citep[\eg][]{lac92,rot94,cha96,blu98,deb00}, and optical ($R-$ or
$I-$band) or $K-$band imaging. Here we discuss the efficiency of our
identification procedure based on the data obtained to date.

It is possible that our relatively low resolution radio maps may have led
to a higher fraction of mis-identifications. We can, however, use some
prior information to determine if the spectroscopically observed object is
the correct host galaxy of the USS radio source. One important tool is the
Hubble $K-z$ diagram (see also \S 5.2.3). \citet{deb02} show that the hosts
of radio galaxies are likely to be about 2 magnitudes brighter than normal
star-forming galaxies at the same redshift.  Coupled to the rarity of
radio galaxies (e.g. \citealp{wil01a}) compared with normal star-forming
galaxies (e.g. \citealp{sey04}), this means that a mis-identification
would most likely be made with a normal galaxy (as opposed to another
radio galaxy host), several  magnitudes fainter than expected for
its redshift. In other words, the faintest near-IR ($K\geq19.5$) galaxies
in our sample are expected to be associated with the hosts of radio
galaxies at redshifts beyond $z\sim1$. A mis-identification of these hosts
would most likely lead to galaxies in the redshift range $0.3\lesssim
z\lesssim 1$. Nine of the ten galaxies we observed with $K\geq19.5$ have
resulted in spectroscopically confirmed redshifts beyond 1.3, from which
we conclude that their identifications are reliable. The tenth turned out
to be at $z=0.826\pm0.002$ and we consider this to be unreliable (see
below). 

\begin{figure}
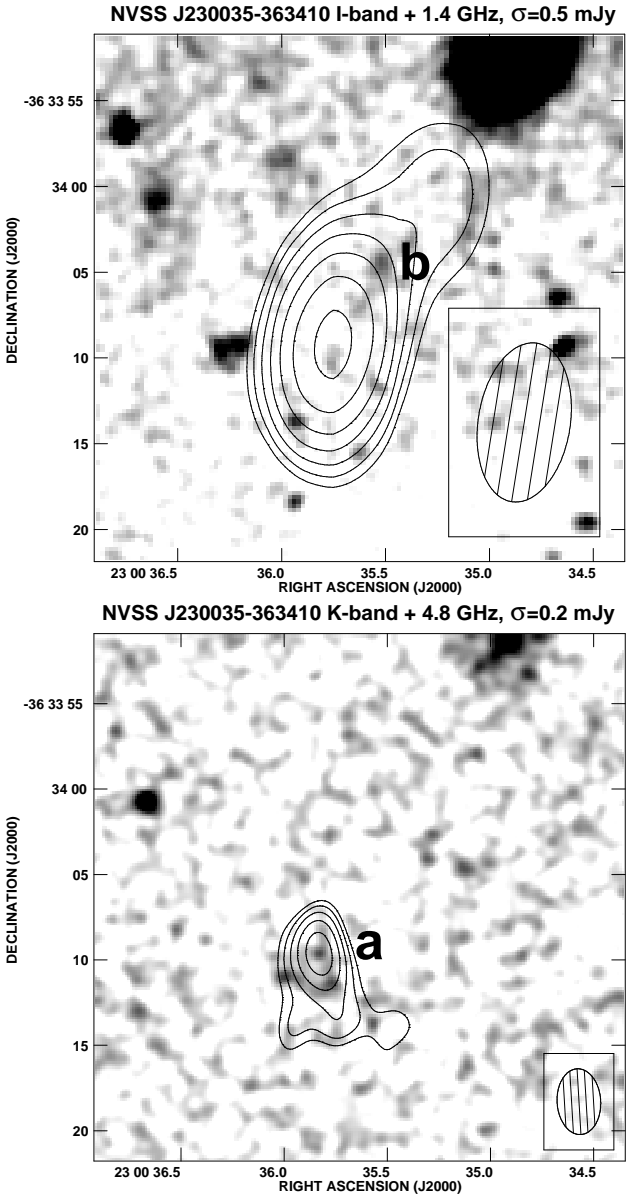
 
\vspace{0.2cm}
\psfig{file=NVSSJ230035-363410I.L.PS,width=8.6cm,angle=-90} 
\vspace{0.2cm}
\psfig{file=NVSSJ230035-363410K.C.PS,width=8.6cm,angle=-90}
\caption{{\it Top:} ATCA 1.4\,GHz image of NVSS~J230035$-$363410 from
paper~I overlaid on a VLT/FORS2 $I-$band image. {\em b} marks the
initial identification. Object~{\em a} is not detected to a limiting
magnitude of $I$=25. {\it Bottom:} ATCA 4.8\,GHz image overlaid on the
NTT/SofI $K-$band image from paper~I. This radio image clearly
identifies object {\em a} as the host galaxy, illustrating the
importance of high-resolution radio maps and $K-$band, rather than
$I-$band imaging. Object~{\em b} is not detected to a limiting
magnitude of $K$=20.6. The contour scheme is a geometric progression in
$\sqrt 2$. The first contour level is at 3$\sigma$, where $\sigma$ is
the rms noise measured around the sources, as indicated above each
plot. The synthesised beams are indicated in the lower right corners.}
\label{NVSS2300K.C} \end{figure}


Of the 37 USS radio sources for which we present spectroscopic
redshifts in this paper, four are clearly misidentified with
foreground objects, with a fifth identification which we regard with
skepticism. As mentioned in \S 3.2, three host galaxy candidates
(NVSS~J231016$-$363624, NVSS~J202856$-$353709 and
NVSS~J231016$-$363624) turned out to be foreground stars. The fourth
misidentification 
was the target
associated with NVSS~J230035$-$363410. In this case, its 20cm radio
morphology (Figure~\ref{NVSS2300K.C},
\textit{top}) is extended, making a clear identification
difficult. Optical imaging revealed an $I=24.6$ galaxy located at
{\bf{b}} on Figure~\ref{NVSS2300K.C} (\textit{top}) to be the most
likely candidate. Follow-up spectroscopy determined the redshift of
this galaxy to be $z=0.826\pm0.002$, based on \OII\ and \OIII\ lines
at $\lambda_{\rm obs}=6806$\,\AA\ and 9146\,\AA, respectively.
According to the best-fitting $K-z$ relation for 3C, 6C and 7C radio
galaxies \citep{wil03}, the expected $K-$band magnitude for a
$z=0.826\pm0.002$ radio galaxy with a radio luminosity typical of a 7C
source is $17.26\pm0.1$. We became skeptical about the reliability of
the identification of NVSS~J230035$-$363410 based on its faint
infra-red magnitude ($K>20$), a $>27\sigma$ deviation from the $K-z$
relation. We have since obtained higher spatial resolution
(3\farcs9$\times$2\farcs6) radio observations at 4.8~GHz and 6.2~GHz
(Fig.~\ref{NVSS2300K.C}{\it , bottom}; Klamer et al, in prep), which
indicate that the position of the radio core is located several
arcseconds south-east of {\bf{b}}, coincident with a $K=19.8$, $I>25$
galaxy labeled {\bf{a}} on Figure~\ref{NVSS2300K.C} ({\it
bottom}). In this case, deep $I-$band imaging was counter-productive
to our identification process, but we were able to recognise the error
by comparing its predicted versus measured location on the $K-z$
diagram.
 
In a similar context, the $z=0.352$ galaxy identified with
NVSS~J011606$-$331241 is much fainter than expected, deviating by
$13\sigma$ from the $K-z$ relation for 7C radio galaxies \citep{wil03}.
Therefore, it is possible we have also misidentified this host galaxy with
a foreground object, although there are examples of similar $K-z$ outliers
with secure host galaxy identifications \citep{wil03}. A close inspection
of the $K-$band image reveals a much fainter object approximately 0\farcs5
to the West. Higher spatial resolution radio observations are necessary in
order to increase the sensitivity and positional accuracy of the radio
emission in order to determine if this fainter source could be the true
host galaxy. Alternatively, the $z=0.352$ galaxy may be a foreground
galaxy acting as a gravitational lens of an as yet undetected background
radio galaxy.  Our NTT $K-$band image (see Fig.~2 in paper~I) does not
show any clear signs of lensing; deeper optical and/or near-IR images with
high spatial resolution would be needed to test this hypothesis.

\section{Discussion}
With spectroscopic observations for 50 out of 73 sources\footnote{We
have excluded the three objects obscured by foreground stars.} (39 out
of 51 if we consider only the sources with $\alpha_{843}^{1400} <
-1.3$) in our sample, we can now start looking at the statistical
properties of our sample. Although we have been able to determine only
35 spectroscopic redshifts, our deep spectra do provide constraints on
the redshifts of the other 15 sources (\S 5.1). Most of our spectra
are of insufficient quality to derive much physical information about
the host galaxies or their extended emission-line regions. However,
with the redshift information, we are now in a position to interpret
their $K-$band fluxes in terms of stellar populations, and compare
them with other samples of radio galaxies (\S 5.2). Finally, the
redshift information for 43/73 = 59\% of the sample, including five
sources at $z>3$, allows us to put strong lower limits on the space
densities of these massive galaxies at high redshift (\S 5.4).

\subsection{The nature of the continuum-only and undetected objects} 
We first discuss the nature of the sources where we detected only
continuum emission (Fig.\ \ref{continua}), or which remained
undetected in medium-deep VLT exposures. The continuum sources have
mainly been observed with the NTT. It is likely that deeper
observations with 8--10m class telescopes may yield a spectroscopic
redshift. The detection of continuum emission down to $\sim$4000\,\AA\
indicates that the \Lya\ discontinuity in these sources must be
bluewards of this, and hence we can constrain their redshifts to
$z\simlt 2.3$. On the other hand, if the \OII\ line was at wavelengths
free of strong night sky lines ($\lambda_{\rm obs}<7200$\,\AA), we
would probably have detected it, so the likely redshift range is
$1<z<2.3$, corresponding to the `redshift desert'. These 8 sources are
thus likely to be high redshift radio galaxies with relatively faint
emission lines, either intrinsically or due to dust obscuration.

Of the six undetected sources, one was observed with the NTT, and may be
detected with more sensitive observations. The remaining five sources are
extremely faint, and are beyond the capabilities of present-day optical
spectrographs. The WN/TN USS sample of \cite{deb01} contains seven such
undetected USS sources in $\sim$1\,hour Keck spectra;  all have unresolved
radio morphologies $\simlt$1\farcs5. Of the five undetected sources in the
SUMSS--NVSS USS sample, four have unresolved radio structures with sizes
$<$6\arcsec, and only NVSS~J015223$-$333833 is resolved: a 15\arcsec\
radio double. They also have some of the steepest radio spectral indices
in our USS sample with $\alpha_{843}^{1400}<-1.55$ (where $S_{\nu} \propto
\nu^{\alpha}$).  This, combined with their faint $K-$band identifications
(two sources have $K>20.5$) suggests these sources may be at $z\simgt7$,
where \Lya\ has shifted out of the optical passband. We have started a
near-IR spectroscopy campaign on these sources with the Gemini-South
telescope to search for emission lines shifted into the $J$ band.

An alternative explanation is that the undetected sources are
heavily obscured by dust. This is likely to be the case in a
significant fraction of the undetected WN/TN USS sources
\citep{deb01}, since at least half show strong (sub)mm thermal dust
emission \citep{reu03,reu05}.  Sub-mm observations of these sources
with the Large Apex Bolometer Camera (LABOCA) on the Atacama
Pathfinder Experiment (APEX) should allow us to determine if they
are indeed strong dust emitters. Optical and near-IR redshift
determination may then prove unfeasible, and alternatives such as
observing the rotational transitions of CO may be needed. Targeted
searches will become possible in the near future with the
wide-bandwidth correlator on the Australia Telescope Compact Array,
and later with the wide bandwidth receivers of the Atacama Large
Millimetre Array (ALMA).

\subsection {The K-z relation}

\begin{figure*}
\centerline{
\psfig{file=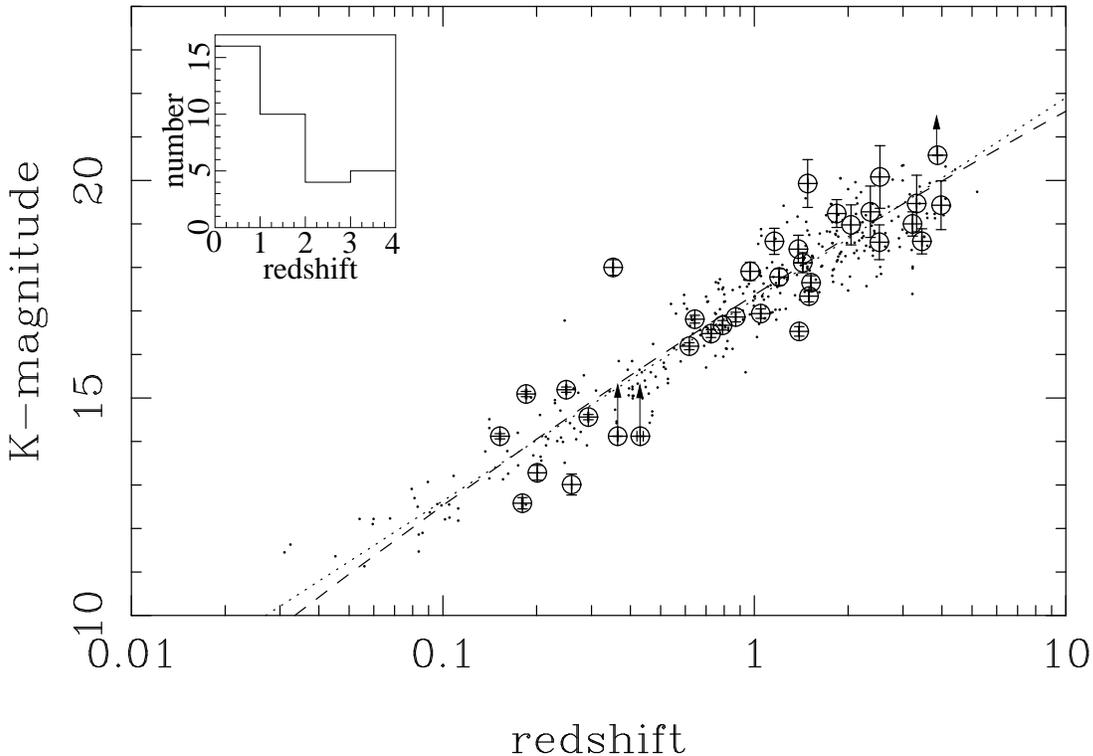,height=10cm}}
\caption{$K-$band magnitude corrected to a 64\,kpc metric aperture versus
redshift. The sources with measured redshifts in
Table~\ref{spectroscopyjournal} are shown by the large circles. In some
cases the $K-$band magnitude uncertainties are smaller than the 
circles, and
the redshift uncertainties are all much smaller than the circles.  The
dots comprise a number of radio catalogues; the 3CRR, 6CE, 6C and 7CRS
radio galaxy samples \citep[compiled by][]{wil03} and the composite
samples of \citet{wvb98} and \citet{deb02}.  The dotted line is the fit to
the galaxy samples (dots) from Paper~I, while the dashed line is the fit
to the galaxy sample from \citet{wil03}.  Inset is the histogram of the
redshifts for the 35 sources measured in this paper.}
\label{K-z}
\end{figure*}

Figure~\ref{K-z} shows the $K-$band magnitudes versus redshifts for
the sources in Table~\ref{spectroscopyjournal}, along with
corresponding data from several radio catalogues listed in the
caption.  We converted the 8\arcsec-diameter aperture magnitudes
from paper~I to 64\,kpc metric apertures using the procedures
described by \citet{eal97}. This consists of using the \citet{san72}
curve of growth for radio galaxies at $z<0.6$ and a radial profile
$\propto r^{0.35}$ for radio galaxies at $z>0.6$.

\subsubsection{Contribution from emission lines}
\label{Kzlines}
The small scatter in the $K-z$ relation is attributed to the stellar
contribution to the $K-$band magnitude. However, at some redshifts
strong emission lines can boost the $K-$band magnitude, offsetting the
distribution towards brighter $K-$band magnitudes.  Our optical
spectroscopy has shown that most of the galaxies have apparently
strong emission lines (see Table \ref{lineparameters}).  However, our
SUMSS--NVSS sample contains sources with radio luminosities which are
almost an order of magnitude lower than those in flux-limited radio
surveys such as 3C, 6C and 7C. The strong correlation between radio
and emission line luminosity \citep{wil99} thus predicts much
weaker emission lines, which will also contribute less to the
broad-band fluxes \citep[see also][]{jar01,roc05}.

The most important emission-line contribution to the $K-$band flux
would come from \Ha, \OII\ and \OIII\ \citep{deb02} at redshifts
around $z\sim2.3$, $3.4$ and $4.9$.  While two of the three measured
points on our $K-z$ plot around $z\sim3.4$, have slightly brighter
$K-$band magnitudes than the fitted line, they are still within the
expected scatter and are therefore not obviously affected by
emission-line contributions. Conversely, at $z\sim2.3$ the scatter
is towards fainter $K-$band magnitudes.  Therefore, there is no
strong evidence in our $K-z$ distribution for contamination of the
$K-$band magnitudes by emission lines.

\subsubsection{Dispersion in the $K-z$ relation}

The dispersion in the $K-$band magnitudes along the $K-z$ diagram
has been used as an indicator of the evolution of the stellar
population in powerful galaxies \citep{eal97,jar01,wil03,roc04}.  
We calculated the dispersion of the measured $K-$band magnitudes
from the line of best fit (Paper~I) and compared this value in
redshift bins.  Galaxies in each unit of redshift from $z=0$--1 up
to $z=3$--4 had standard deviations of 1.1, 1.0, 0.7 and 0.8
respectively. The standard deviation for $z<2$ was 1.0, decreasing
to 0.7 for $z>2$.

\citet{eal97} found an increase in the dispersion of $K-$band
magnitude for $z>2$ galaxies and used this to support their model
that $z>2$ is the epoch of galaxy formation.
Our results do not support this model, but instead we see a slight
decrease in dispersion at higher redshifts, which is more consistent
with the observations of \citet{jar01}. We need to caution that if
there is a significant contribution to the $K-$band magnitude from
non-stellar emission (scattered or transmitted AGN light, emission
lines), then the assumption that the dispersion in the $K-z$ plot is
indicative of the stellar evolution is compromised.

While redshifts have not been measured for our complete sample, the
low scatter around the $K-z$ relation seems at least consistent with a
model in which the epoch of formation of ellipticals may be at a
similar high redshift ($z>4$), with passive evolution since then.  At
$z>1$, the SUMSS--NVSS sources also appear slightly fainter than the
3C, 6C and 7C radio galaxies.  If the $K-$band light is indeed
dominated by stars, this can be interpreted as a lower average mass in
the range 10$^{11}$--10$^{12}$\,M$_{\odot}$
\citep{roc04}. This would be consistent with the lower radio
luminosities of the SUMSS--NVSS sources implying less massive central
black holes in their host galaxies.

\subsubsection {Accuracy of the $K-z$ relation as a redshift indicator}
In Paper~I we used the $K-z$ relation to predict the median redshift
of our sample to be 1.75. The distribution of measured redshifts for
our sample is shown in Fig.~\ref{K-z}, with a median $z$ of 1.2.  If
the galaxies listed as `undetected' in Table~\ref{spectroscopyjournal}
are not detected because they are at high redshift (at least
$z>1.75$), then this would shift the median $z$ up to 1.5. There also
remains a further 23 sources for which the redshift is yet to be
determined; 15 of the 23 are faint enough ($K[8\arcsec]>18.3$) to lie
at $z>1.75$, therefore increasing the median $z$ to at least 1.75.
Therefore we anticipate that the median redshift will be close to that
predicted when we have spectroscopic redshifts for the complete
sample.

The redshift distribution from the $K-z$ relation included at least
three galaxies with $z>4$. Follow-up spectroscopy has yet to find
any galaxies with $z>4$. The best fit to the radio galaxy $K-z$
relation predicts $z>4$ for $K>20.06$ host galaxies.  Of the four
galaxies in our sample that have $K(\rm 64\,kpc)>20.06$, two have
yet to be observed spectroscopically, and the remaining two have
redshifts measured to be 2.531 and 1.483.  The latter galaxy is on
the extremities of the $K-$band magnitude distribution for radio
galaxies, and while the former target is within the scatter, it is
far from the predicted $z=4$.  Conversely, a target predicted from
the K-magnitude, to be at $z=2.15$ turned out to be at $z=3.450$.
This highlights one of the shortcomings of the $K-z$ plot, namely
that there is a broad range of redshifts for any given $K-$band
magnitude. Despite this, the $K-z$ relation remains a useful tool
for predicting redshift when planning follow-up spectroscopy.

\subsection{Surface density of SUMSS--NVSS USS radio galaxies at $z>3$ }

As noted in paper~I, the surface density of USS radio sources selected
from NVSS and SUMSS (482\,sr$^{-1}$) is more than four times higher
than that of USS sources selected by \citet{deb00,deb02} from the
WENSS and NVSS surveys (103\,sr$^{-1}$ to the same 1.4\,GHz flux
density limit of 15\,mJy). We now ask whether the higher surface
density of SUMSS--NVSS objects translates into a correspondingly
higher surface density of high-redshift radio galaxies. Table~4
compares the redshift distribution of USS objects from the two
surveys.  The fraction of $z>3$ objects in the SUMSS--NVSS sample is
similar to that found by \citet{deb00,deb02}, implying that USS
selection at 843--1400\,MHz is also an efficient way of selecting
$z>3$ radio galaxies.

\begin{center}
\begin{table}
\caption{Redshift distribution of USS sources 
from this sample and \citet{deb01}.  Objects with spectral 
index $\alpha>-1.3$ have
been excluded from the statistics.  }
\begin{tabular}{@{}lrcrc}
\hline
           & \multicolumn{2}{l}{This paper} & \multicolumn{2}{l}{De 
Breuck et al.\ (2001)} \\
\hline
$z<3$      &   23 & $59\pm12$\%& 37 & $61\pm10$\% \\
$z>3$      &    5 & $13\pm6$\% & 11 & $18\pm5$\% \\
Continuum  &    5 & $13\pm6$\% &  6 & $10\pm4$\% \\
Undetected &    6 & $14\pm6$\% &  7 & $11\pm4$\% \\
Total      &   39 &            & 61 &              \\
\hline
\end{tabular}
\noindent
\end{table}
\end{center}

We can use the results in Table 4, together with the known surface density
of USS radio sources, to estimate a minimum surface density for powerful
radio galaxies at $z>3$.  If $13\pm6$\% of SUMSS--NVSS USS radio sources
lie at $z>3$, then the minimum surface density of $z>3$ radio galaxies is
63$\pm$29\,sr$^{-1}$.  This is higher than the corresponding value for the
WENSS--NVSS USS sample (19$\pm$5\,sr$^{-1}$), implying that the
SUMSS--NVSS selection method can identify about three times as many
genuine $z>3$ objects per unit area of sky as the WENSS--NVSS selection.

\subsection{Space density of radio galaxies at $3<z<4$ }

For $3<z<4$, the 1.4\,GHz flux limit of the SUMSS--NVSS sample
corresponds to a minimum radio power of roughly 10$^{27.0}$ to
10$^{27.2}$\,W/Hz.  For USS radio galaxies above this luminosity,
using the surface densities derived above, we can estimate a minimum
space density of roughly $1.2\pm0.6\times10^{-9}$\,Mpc$^{-3}$. This is
very close to the density of powerful steep-spectrum radio galaxies
predicted at $3<z<4$ by \citet{dun90}, using models based on a
complete sample of radio sources selected at 2.7\,GHz.  Their models
show a gradual (but modest) decline in the space density of powerful
radio galaxies over the range $z\sim2-4$.  Our current data are
consistent with these models if USS radio galaxies represent the
majority of powerful radio galaxies at $z>3$.  If there are also
significant numbers of $z>3$ radio galaxies with $\alpha>-1.3$ (which
would not be detected in our current survey), then the space density
of powerful radio galaxies at these redshifts would be higher than
predicted by the Dunlop \& Peacock models.

\section{Conclusions}

Based on optical spectroscopy of 53/76 of the sources in the SUMSS--NVSS
USS sample, we draw the following conclusions:

\begin{itemize}

\item We obtain 35 spectroscopic redshifts, including five radio galaxies
at $z>3$.  We also found three quasars at 1$< z <$1.6.

\item Seven sources show only continuum emission, with no clear emission or
absorption lines. These are probably sources at $1\simlt z \simlt2.3$,
which are either intrinsically faint or obscured by dust.

\item Eight sources remain undetected in 0.5--2.25\,hour deep VLT
spectra. These sources could be either heavily obscured by dust, or they
could be at $z \simgt 7$.

\item We have obtained $I-$band imaging of 10 USS sources down to
$I\approx 25$; this turned out to be counter-productive in identifying the
host galaxies. Medium deep $K-$band imaging combined with high-resolution
($<$5\arcsec) radio imaging is a more efficient identification procedure.

\item The SUMSS--NVSS radio galaxies generally follow the $K-z$ relation
defined by other radio galaxies, with one notable exception, which may be
a gravitationally amplified object. The dispersion about the $K-z$
relation slightly decreases at $z>2$, contrary to the results seen in the
6CE sample \citep{eal97}, and more consistent with the 6C$^*$ USS sample
\citep{jar01}.

\item We derive a strict lower limit on the space density of $3<z<4$
radio galaxies of $1.2\pm0.6\times10^{-9}$\,Mpc$^{-3}$.

\end{itemize}

The five new $z>3$ objects discovered from this sample bring the total
number of known $z>3$ radio galaxies in the southern hemisphere from 7 to
12, compared to 19 in the northern hemisphere. These new sources now
provide sufficient targets for detailed follow-up studies with large
optical and (sub)millimetre facilities in the southern hemisphere.

\section{Acknowledgements}
This work was supported by PICS/CNRS (France) and IREX/ARC (Australia).

{}

\end{document}